\documentclass[%
 reprint,
 amsmath,amssymb,amsfonts,
 aps,
]{revtex4-2}

\usepackage[dvipdfmx]{graphicx}
\usepackage{tabularx}
\usepackage{dcolumn}
\usepackage{bm}
\usepackage{color}
\usepackage{ulem}
\usepackage{soul}
\usepackage{multirow}
\usepackage{mathtools}
\newcounter{figcountSI}

\newcounter{seccountSI}

\def\mat#1{#1}
\def\stat#1{#1}

\def\ket#1{\mbox{\boldmath $#1$}}

\def\nHr#1#2{\ensuremath{\left(\kern-.3em\left(\genfrac{}{}{0pt}{}{#1}{#2}\right)\kern-.3em\right)}}

\newcommand{\argmin}{\mathop{\rm arg~min}\limits}
\newcommand{\Perm}{P}
\newcommand{\ER}{Erd\H{o}s-R\'{e}nyi}
\newcommand{\SBM}{stochastic block model}

\newcommand{\Methods}{Appendix}

\setstcolor{red}

\begin{document}


\title{Sequential locality of graphs and its hypothesis testing}


\author{Tatsuro Kawamoto}
\affiliation{Artificial Intelligence Research Center, \\
  National Institute of Advanced Industrial Science and Technology, 
  Tokyo, Japan }

\author{Teruyoshi Kobayashi}
\affiliation{Department of Economics, Kobe University, Hyogo, Japan}


\date{\today}

\begin{abstract}
The adjacency matrix is the most fundamental and intuitive object in graph analysis that is useful not only mathematically but also for visualizing the structures of graphs. 
Because the appearance of an adjacency matrix is critically affected by the ordering of rows and columns, or vertex ordering, statistical assessment of graphs together with their vertex sequences is important in identifying the characteristic structures of graphs. 
In this paper, we propose a hypothesis testing framework that assesses how locally vertices are connected to each other along a specified vertex sequence, which provides a statistical foundation for an optimization problem called envelope reduction or minimum linear arrangement. 
The proposed tests are particularly suitable for moderately small data and formulated based on a combinatorial approach and a block model with intrinsic vertex ordering. 
\end{abstract}

\maketitle

\section{Introduction \label{sec:Introduction}}
Much effort has been devoted to identifying the characteristic structures in graph data~\cite{barabasi2012takeover,newman2018networks}. 
The most fundamental representation of a graph is the adjacency matrix, in which the row and column indices correspond to the vertices, and the matrix elements represent the connectivity among the vertices. 
Adjacency matrix is not only essential in theoretical graph analysis, but also useful for visualization \cite{Siirtola99,FriendlyKwan2003,Wu2008,Perin2014}. 

\begin{figure*}[ht!]
  \centering
  \includegraphics[width= 1.99\columnwidth]{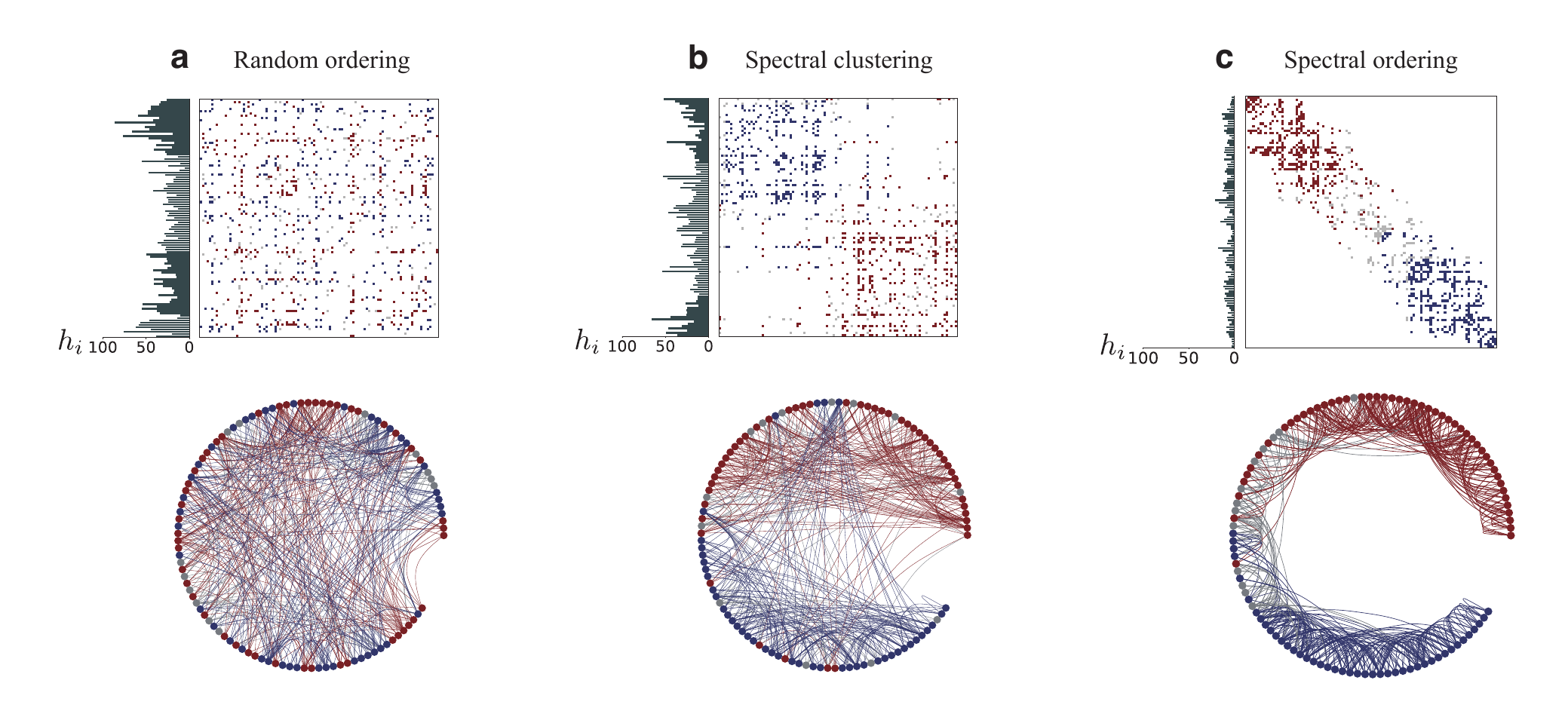}
  \caption{
  The \textit{political books} dataset with different orderings of the adjacency matrix elements. 
  In (a), the vertices are ordered randomly. 
  In (b) and (c), the orderings are determined by the spectral clustering (i.e., community detection) and spectral ordering, respectively; in both cases, we use the normalized Laplacian as the regularized matrix. 
  This graph has metadata on each vertex, showing the category of each book: We indicate them in red (``conservative''), blue (``liberal'), and gray (``neutral'') in the network plots. 
The adjacency matrix elements and the edges in the network plot are indicated in red/blue if both ends of the edge are ``conservative''/``liberal''; otherwise, they are indicated in gray. 
  In each panel, the bar plot represents the microscopic measure of locality $h_{i}\coloneqq {\rm median}\{|\pi_i - \pi_j |: A_{ij}>0, j\in I\}$ for $i \in I$.}
  \label{fig:Polbooks}
\end{figure*}

However, the appearance of an adjacency matrix critically depends on vertex ordering, and the interpretation of the graph structure can differ as vertex ordering varies.  This issue has often been ignored, mainly because vertices usually do not have intrinsic ordering, and it is common to study characteristics that are invariant under a permutation of vertex labels.
Figure~\ref{fig:Polbooks} shows adjacency matrices of the same graph, called \textit{political books} \cite{Newman2006politicalbooks}, with different vertex orderings 
(the vertices represent books about US politics, and two vertices are connected if the two books are co-purchased frequently). 
When the vertices are ordered randomly, as shown in Fig.~\ref{fig:Polbooks}{\bf a}, the adjacency matrix is apparently uniformly random. 
On the other hand, when we run a community detection algorithm (e.g., spectral clustering \cite{ShiMalik2000,LuxburgTutorial}) and align the vertices such that the vertices with the same group label are close to each other, as shown in Fig.~\ref{fig:Polbooks}{\bf b}, we can identify a nearly block-diagonal structure, indicating that the graph consists of densely connected components, a structure usually referred to as community structure. 
As the \textit{political books} dataset has a label for each vertex (``conservative,'' ``liberal,'' and ``neutral''), it is confirmed that the two ``blocks'' at the corners of the adjacency matrix are associated with the sets of ``conservative'' and ``liberal'' vertices. 
However, this is not the only structure that can be found in this dataset. 
If we optimize the ordering so that the nonzero elements are concentrated around the diagonal components, we obtain the adjacency matrix shown in Fig.~\ref{fig:Polbooks}{\bf c}. 
In other words, the vertex sequence is permuted so that the sum of distances between connected vertices along the sequence is minimized (we used a method called spectral ordering \cite{DingHe2004} in Fig.~\ref{fig:Polbooks}{\bf c}). 
Such an optimization is known as \textit{envelope reduction}~\cite{Barnard95,DingHe2004} or \textit{minimum linear arrangement} \cite{Harper1964,Chung1984,RaoRicha2004,Devanur2006,Seitz2010}, which is also closely related to the seriation problem \cite{Liiv2010,Behrisch2016}, the consecutive ones problem (C1P) \cite{Robinson1951,Fulkerson1965,Kendall1969,Atkins98,Vuokko_SIAM2010}, and the $k$-sum minimization problem \cite{Fogel_NIPS2013}. 
The optimized adjacency matrix reveals that, in addition to the community structure, the \textit{political books} dataset also has a locality structure along the optimized sequence.

In this paper, we establish a statistical framework for the envelope reduction problem, which is formulated as an optimization problem that aims to find the best vertex ordering. 
While optimization algorithms and their efficiencies have been studied extensively in computer science, the extent to which such optimized ordering reflects the inherent structure of a graph is generally unknown. 
Therefore, independently of finding the best vertex ordering, a method that enables us to assess whether the adjacency matrix with a given vertex sequence exhibits a statistically significant structure needs to be developed. 

To this end, we introduce a notion of \textit{sequential locality}. 
It is a test statistic that captures how locally vertices are connected to each other along a given sequence based on a given metric. 
We let $G = (V, E)$ be a graph, where $V$ ($|V| = N$) denotes the set of vertices and $E$ ($|E| = M$) is the set of edges. 
We refer to the set of raw (or original) vertex indices as $I = \{1, \dots, N\}$ and define the inferred sequence $\ket{\pi} = \{ \pi_{i} | i \in I, \pi_{i} \in I \}$ as a permutation of the raw indices. 
We let $\mat{A}$ be the adjacency matrix for a graph, where $A_{ij}$ represents the number of edges between the $i$th and $j$th vertices in the original sequence. 
Throughout this paper, we consider undirected graphs without self-loops. 
We consider both simple graphs and multigraphs i.e., graphs with multiedges (multiple edges between a pair of vertices). 
We specify a metric that represents the affinity among vertices by a matrix $\mat{J}$.
Then, the measure of sequential locality is defined as 
\begin{align}
\stat{H}_{J} 
= \sum_{i<j} A_{ij} J_{\pi_{i} \pi_{j}}, \label{GeneralSequentialLocality}
\end{align}
where $\sum_{i<j}$ is the sum with respect to $i \in I$ and $j \in I$ with $i<j$.

We employ $\stat{H}_{J}$ as a test statistic and use it to assess the statistical significance of graphs and their inferred vertex sequences. 
$\stat{H}_{J}$ is a quantity that depends on graph $\mat{A}$, vertex sequence $\ket{\pi}$, and affinity matrix $\mat{J}$. 
Traditionally, the envelope reduction problem employs the squared Euclidean distance $J_{ij} \propto (i-j)^{2}$, and the minimum linear arrangement problem employs the sequential distance $J_{ij} \propto |i-j|$ as the affinity matrix. 
Alternatively, we can also consider other metrics such as the logarithmic semimetric $J_{ij} \propto -\log (1 - |i-j|/N)$, or a piece-wise constant distance $J_{ij} \propto \theta(|i-j| - r)$, where $r$ is a constant and $\theta(\cdot)$ is the step function. 
In these cases, $\stat{H}_{J}$ indicates the degree of non-locality. 
In the current study, we employ the sequential distance as $\mat{J}$ (see Refs.~\cite{Wu2008,Hahsler2017,Morone2021} for other measures considered in the literature). 

Equation (\ref{GeneralSequentialLocality}) can also be expressed as 
\begin{align}
\stat{H}_{J} 
= \frac{1}{2}\mathrm{tr} \mat{A} \left( \mat{\Perm}_{\pi} \mat{J} \mat{\Perm}^{\top}_{\pi} \right) 
= \frac{1}{2}\mathrm{tr} \left( \mat{\Perm}^{\top}_{\pi} \mat{A} \mat{\Perm}_{\pi} \right) \mat{J}, 
\label{GeneralSequentialLocality2}
\end{align}
where $\mathrm{tr}$ represents the trace operation and $\mat{\Perm}_{\pi}$ and $\mat{\Perm}^{\top}_{\pi}$ are the permutation matrix and its transpose. 
Thus Eq.~(\ref{GeneralSequentialLocality}) can be interpreted as a quantity obtained after permuting $\mat{A}$ to $\mat{\Perm}^{\top}_{\pi} \mat{A} \mat{\Perm}_{\pi}$ for a given $\mat{J}$. 
Although $\mat{A}$ and $\mat{J}$ apparently play symmetrical roles, they are conceptually distinct objects; $\mat{J}$ defines the similarity between each pair of vertices regardless of the graph structure, while the adjacency matrix $\mat{A}$ is responsible for vertex connectivity in a graph. 
A possible generalization of Eq.~(\ref{GeneralSequentialLocality}) would be to replace $\mat{A}$ with a modified matrix $\mat{A}^{\prime}$, whose element represents the shortest-path distance \cite{Behrisch2016,kwon2021deep} between a pair of vertices, although such an extension is beyond the scope of this paper. 

We can also consider a microscopic measure of the sequential locality for each vertex. 
For example, among the sequential distances between neighboring vertices, we can use their maximum or median as the degree of sequential locality of a target vertex. 
In Fig.~\ref{fig:Polbooks}, we plot the microscopic sequential localities based on medians, which we denote by $h_{i}$ ($i \in I$), as a bar plot in each panel. 
While $h_i$ may not be suitable for a statistical assessment of the entire graph, it is a useful measure for quantifying the degree of locality of each vertex.

We define that a graph exhibits significant sequential locality if there exists a vertex sequence such that $\stat{H}_{J}$ is significantly small compared with those realized under a null hypothesis. 
We refer to a sequence obtained as a solution of an envelope reduction algorithm as an optimized sequence, regardless of its statistical significance.
In the following analysis, we first assess the statistical significance of sequential locality based on an unoptimized vertex sequence. 
Next, we develop statistical tests for optimized sequences. 
Finally, we formulate a statistical assessment under the null hypothesis that the vertex sequence is randomly ordered.

\section{Sequential locality of unoptimized sequences \label{sec:SequentialLocalityUnoptimized}}

\subsection{Statistical test \label{sec:UnoptimizedTest}}
We consider a statistical test for graphs with a given vertex sequence $\ket{\pi}$ that is not explicitly optimized to achieve a small value of $\stat{H}_{J}$. 
In other words, vertices are ``naturally'' ordered (e.g., the original indexing in the dataset). 
To assess statistical significance, we evaluate whether the observed sequential locality can be commonly achieved by the graphs generated by a uniform random graph model. 

We denote $\stat{H}_{1}(\mat{A}, \ket{\pi})$ as the $\stat{H}_{J}$ test statistic with the sequential distance. 
Specifically, 
\begin{align}
\stat{H}_{1}(\mat{A}, \ket{\pi}) 
= \frac{1}{\beta_{1}}\sum_{i<j} A_{ij} \left|\pi_{i} - \pi_{j}\right|, \label{H1_Definition}
\end{align}
where $\beta_{1} = M(N+1)/3$ is a normalization factor. 
As shown below, $\beta_1$ corresponds to the mean value of $\sum_{i<j} A_{ij} \left|\pi_{i} - \pi_{j}\right|$ under a uniform random graph model. 
In fact, $\stat{H}_{1}$ is equivalent to the objective function considered in the minimum linear arrangement problem \cite{Harper1964}, and its minimization is known to be NP-complete \cite{Garey1976}. 
Note, however, that our objective here is to provide a statistical test for a given vertex sequence, not to solve an optimization problem. 

As a uniform random graph model, we consider the Erd\H{o}s-R\'{e}nyi random graph model ({\ER} model, henceforth) with a fixed number of edges, allowing multiedges. 
A graph instance is generated uniformly randomly from all possible graphs with $N$ vertices and $M$ edges. 
Because every element in the adjacency matrix is statistically identical, we consider the following random variable that approximately obeys the distribution for $\stat{H}_{1}(\mat{A}, \ket{\pi})$: 
\begin{align}
\mathsf{H}_{1} = \frac{1}{\beta_{1}} \sum_{m=1}^{M} \mathsf{X}_{m}, \label{H1IID}
\end{align}
where $\mathsf{X}_{m} \in \mathbb{N}$ is a random nonnegative integer drawn from the discrete triangular distribution: 
\begin{align}
\mathrm{Prob}\left[ \mathsf{X}_{m} = x \right] = 
\begin{cases}
\frac{2(N - x)}{N(N-1)} & (0 < x \le N-1) \\
0 & (\text{otherwise})
\end{cases}. \label{TriangularDistribution}
\end{align}
This is because the number of elements with $\left|\pi_{i} - \pi_{j}\right| = x$ is $N-x$ in the affinity matrix, which determines the frequency of the outcome of $\mathsf{X}_{m}$. 
Therefore, we have 
\begin{align}
& \mathrm{Prob}\left[ \stat{H}_{1}(\mat{A}, \ket{\pi}) = E \right] 
\simeq \mathrm{Prob}\left[ \mathsf{H}_{1} = E \right] \notag\\
&= \sum_{\{ 1 \le x_{m} \le N-1\}} \delta\left( \beta_{1} E, \sum_{m=1}^{M} x_{m} \right) 
\prod_{m=1}^{M} \frac{2 (N - x_{m})}{N(N-1)} \label{H1NullDistributionExact}
\end{align}
as the null probability, where $\delta(a,b)$ is the Kronecker delta. 
The normalization factor $\beta_{1}$ in Eq.~(\ref{H1_Definition}) is determined from the fact that $\mathbb{E}\left[ \mathsf{X}_{m} \right] = (N+1)/3$. 

In the limit of large $M$, the central limit theorem guarantees that $\mathsf{H}_{1}$ asymptotically follows a normal distribution. 
Thus, 
\begin{align}
\mathrm{Prob}\left[ \sqrt{\frac{2M (N+1)}{N-2}} (\mathsf{H}_{1} - 1) \le a \right] 
= \int_{-\infty}^{a} \frac{dx}{\sqrt{2\pi}} e^{-\frac{1}{2}x^{2}}. \label{H1NullDistributionNormal}
\end{align}
In fact, this is a moderately accurate estimate of the distribution even when $M$ is not very large, as long as a graph is sparse. 
Hereafter, we denote the standardized $\stat{H}_{1}$ statistic, or the z-statistic as 
\begin{align}
z_{1}(\mat{A}, \ket{\pi}) = \sqrt{\frac{2M (N+1)}{N-2}} (\stat{H}_{1}(\mat{A}, \ket{\pi}) - 1). \label{H1_to_z1}
\end{align}
The standardization factors are obtained by calculating the mean and variance of $\mathsf{H}_{1}$, which are derived in {\Methods}~\ref{sec:ERrandom-H1-Appendix}. 

\begin{figure}[t]
  \centering
  \includegraphics[width= 0.99 \columnwidth]{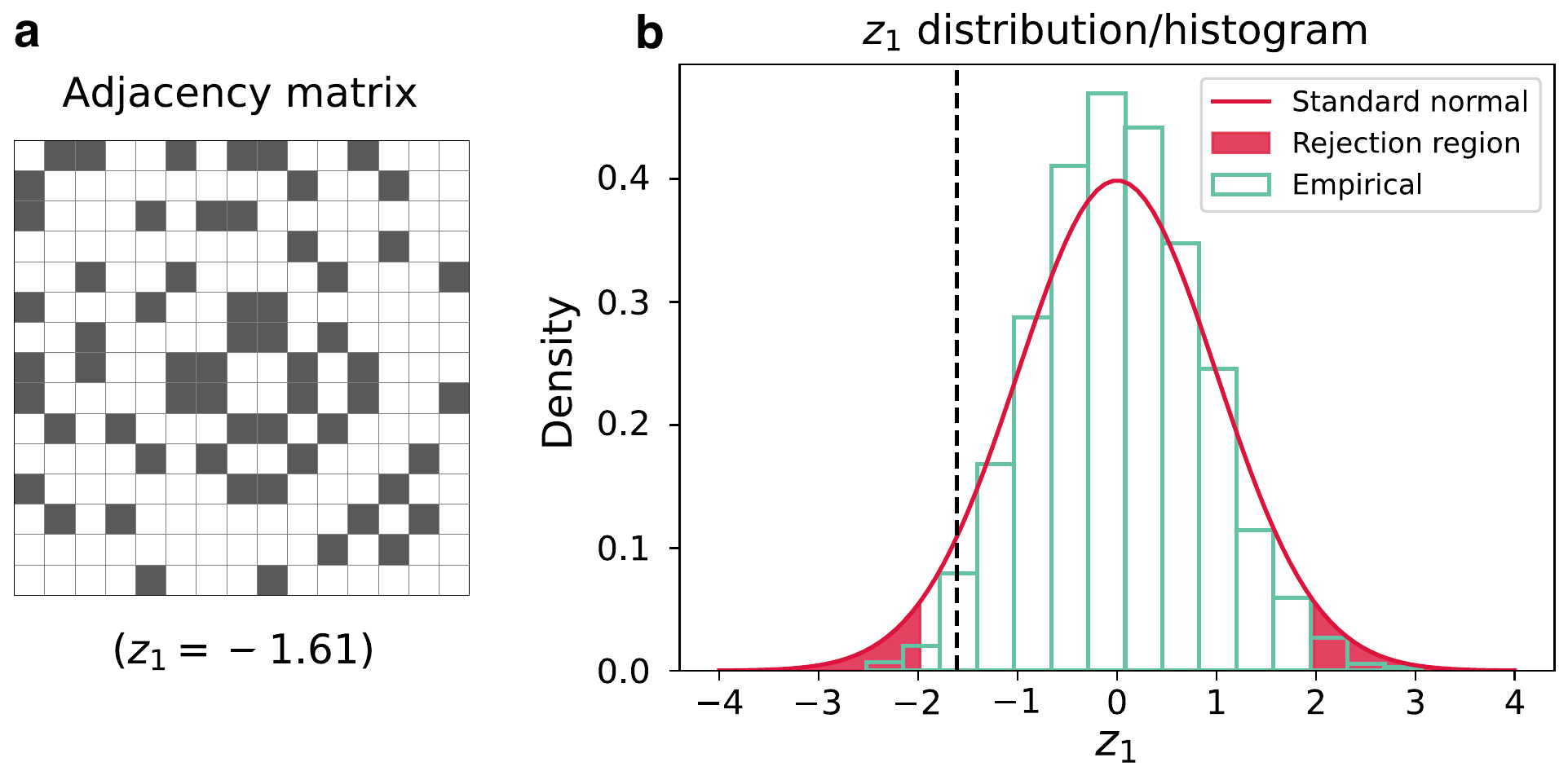}
  \caption{
  Hypothesis testing of an adjacency matrix with unoptimized vertex sequence.
  (a) An adjacency matrix of a small graph (white cells for $A_{ij}=0$ and black cells $A_{ij}=1$) that has $z_{1}=-1.61$. 
  (b) The null hypothesis is not rejected (with $\alpha = 0.05$) because the $z_{1}$ value (dashed line) is not located in the rejection region. 
  The empirical distribution (green histogram) is obtained by calculating $z_{1}$ values for graphs generated from the {\ER} model without multiedges. 
	}
  \label{fig:H1_test_example}
\end{figure}

Using Eq.~(\ref{H1NullDistributionNormal}), we can conduct the significance test of sequential locality (see Fig.~\ref{fig:H1_test_example} for an example). 
Under the null hypothesis that the graph is generated from the {\ER} model, we reject the hypothesis if 
\begin{align}
|z_{1}(\mat{A}, \ket{\pi})| > z^{\ast}_{1}(\alpha), \label{Ztest}
\end{align}
for a given significance level $\alpha$. 
Here, $z^{\ast}_{1}(\alpha)$ is defined such that $\Phi(z < -z^{\ast}_{1}(\alpha)) = \alpha/2$, where $\Phi$ is the standard normal cumulative distribution function. 
Note that Eq.~(\ref{Ztest}) represents a two-sided test that allows us to make a dichotomous decision as to whether we can reject the hypothesis that the graph is generated from the {\ER} model or not. 
In other words, because no model is assumed as an alternative hypothesis, a low value of $z_{1}(\mat{A}, \ket{\pi})$ being smaller than $-z^{\ast}_{1} (\alpha)$ does not necessarily imply that the graph is generated from a model that typically yields a stronger sequential locality (i.e., smaller $z_{1}(\mat{A}, \ket{\pi})$) than the {\ER} model. 
We can only conclude that the observed graph \textit{happened to} exhibit a strong sequential locality when $z_{1}(\mat{A}, \ket{\pi})$ is relatively small.

In passing, we discuss how the entropic evaluation in Eq.~(\ref{H1NullDistributionExact}) differs from the exact distribution. 
Note that the order of the outcome of the sequence $(\mathsf{X}_{1}, \dots, \mathsf{X}_{M})$ matters in Eq.~(\ref{H1_Definition}). 
It implies that the same graphs with different edge-orderings are overcounted. 
This overcounting would have no effect on the distribution if all graph instances are simple, because every graph is overcounted exactly $M!$ times. 
However, the order of the edges within a multiedge is not distinguished \cite{Peixoto2012,Fosdick2018}. 
In other words, when every outcome of $(\mathsf{X}_{1}, \dots, \mathsf{X}_{M})$ is reweighted by $1/M!$, the contribution from the multigraphs is counted less than it should be (see {\Methods}~\ref{sec:XmStatistic} for an illustration using a small graph). 
Therefore, Eq.~(\ref{H1NullDistributionExact}) can be regarded as the distribution for the {\ER} model that is mildly restricted to simple graphs. 

In {\Methods}~\ref{sec:ExactERrandom-Appendix}, we show the exact distribution of a test statistic in which the multiedges are counted correctly and discuss the relationship with the present result. 
However, it should be emphasized that the exact distribution is not necessarily a better choice. 
The assessment using Eq.~(\ref{H1NullDistributionNormal}) is more appropriate than that based on the exact distribution when simple graphs are assumed as the null hypothesis. 
Moreover, we cannot apply the central limit theorem to the exact distribution, implying that we cannot evaluate $p$ values efficiently. 
 
Although we consider a model in which every graph instance has exactly $M$ edges (i.e., the microcanonical constraint), we could alternatively consider the {\ER} model in which the number of edges is constrained only on average (i.e., the canonical constraint) as a null hypothesis.
In fact, its exact test-statistic distribution asymptotically coincides with Eq.~(\ref{H1NullDistributionNormal}) as $M$ becomes large (see {\Methods}~\ref{sec:CanonicalERrandom-Appendix}).

\subsection{Ordered random graph model and the power analysis \label{sec:ORGM}}
To analyze the performance of the statistical test using Eq.~(\ref{H1NullDistributionNormal}), we introduce a random graph model that has an intrinsic vertex sequence exhibiting a desired strength of sequential locality. 
That is, edges are generated with high probabilities between vertices that are deemed to be close to each other in the intrinsic sequence. 
We refer to this model as the \textit{ordered random graph model} (ORGM). 
This model is categorized in the family of block models; as we describe below, the ORGM partly overlaps with the {\SBM} \cite{holland1983stochastic,WangWong87,Peixoto2012}. 
Note that there are several distance-dependent random graph models that have been proposed in the literature, such as the latent space model \cite{Hoff2002,Handcock2007}, geometric random graphs \cite{Penrose2003}, and some random graph models \cite{Grindrod2002,SongWang2014} inspired by the Watts-Strogatz model \cite{WattsStrogatz1998}. 

\begin{figure}
  \centering
  \includegraphics[width= 0.85\columnwidth]{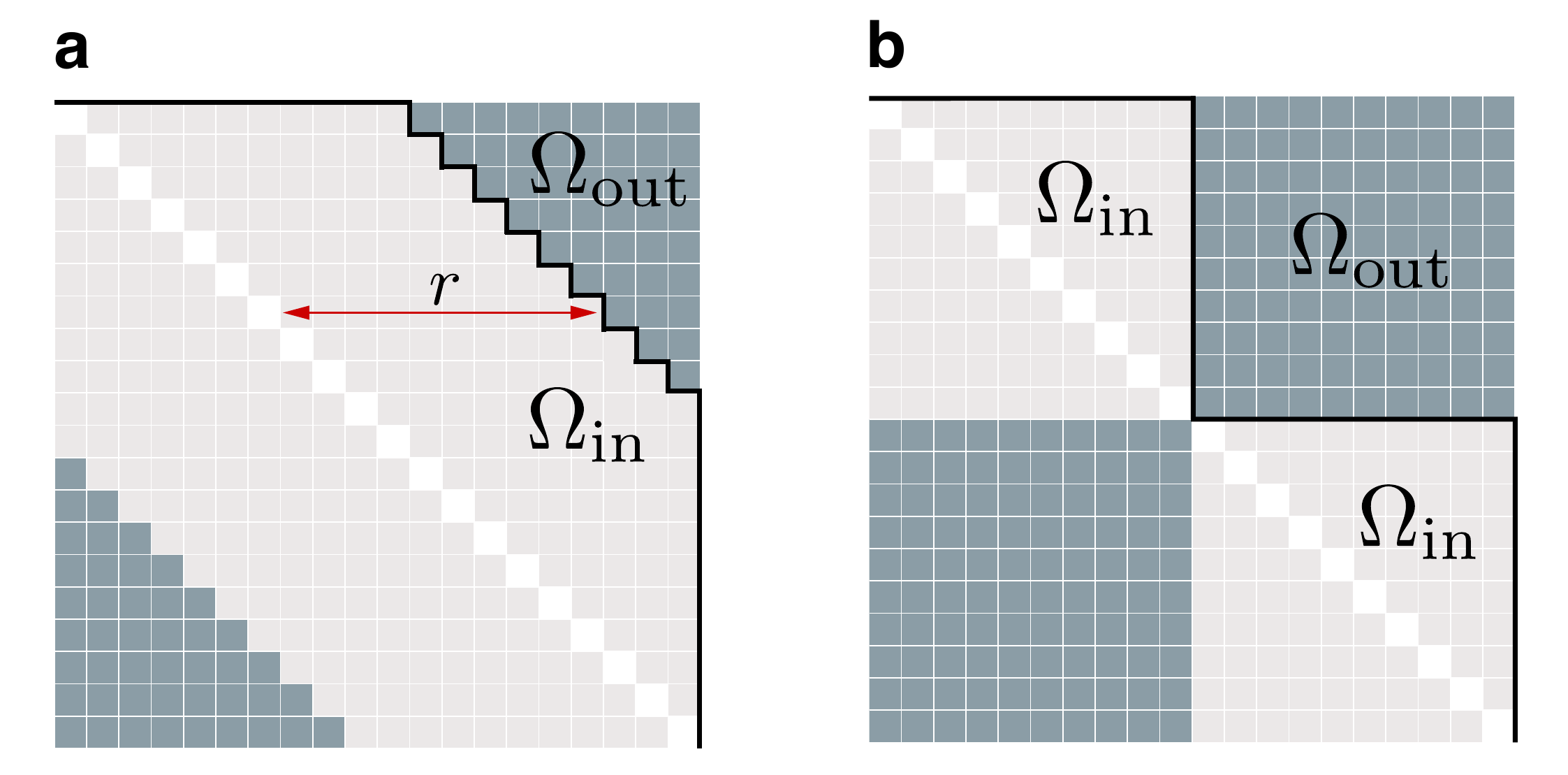}
  \caption{
ORGM with (a) banded and (b) community structures.
  We randomly select $M_{\mathrm{in}}$ elements within $\Omega_{\mathrm{in}}$ (light-shaded cells) with or without repetition. 
  We randomly select $M_{\mathrm{out}}$ elements within $\Omega_{\mathrm{out}}$ (dark shaded cells) with or without repetition. 
  The boundary between the dark- and light-shaded regions (solid line) in each panel represents the envelope function $F(i)$. 
	}
  \label{fig:ORGM_adjacency_matrix}
\end{figure}

We define an envelope function $F(i) \in I$, which is a discrete function of vertex index that specifies the upper bound of the sequential distance below which each pair of vertices are regarded as being close to each other. 
As illustrated in Fig.~\ref{fig:ORGM_adjacency_matrix}, we denote $\Omega_{\mathrm{in}}$ as the set of the upper-right triangle elements of the adjacency matrix that satisfies $|i-j| \le F(i)$ ($i,j \in I$). 
The rest of the upper-right triangle elements is denoted by $\Omega_{\mathrm{out}}$. 
We randomly draw $M_{\mathrm{in}}$ edges for vertex pairs belonging to $\Omega_{\mathrm{in}}$. 
Similarly, we randomly draw $M_{\mathrm{out}}$ edges for vertex pairs belonging to $\Omega_{\mathrm{out}}$. 
The probability distribution of the adjacency matrix $\mat{A}$ for the ORGM is given by 
\begin{align}
\mathrm{Prob}\left[ \mat{A} \right] 
&= \frac{1}{\mathcal{N}_{G}} 
\prod_{i=1}^{N} \delta\left( A_{ii}, 0 \right)
\prod_{i<j} \delta\left( A_{ij}, A_{ji} \right) \notag\\
&\times 
\delta\left( M_{\mathrm{in}}, \sum_{(i,j) \in \Omega_{\mathrm{in}}} A_{ij} \right) 
\delta\left( M_{\mathrm{out}}, \sum_{(i,j) \in \Omega_{\mathrm{out}}} A_{ij} \right), \label{ORGM-Prob}
\end{align}
$\mathcal{N}_{G}$ is the total number of graphs, which can take different values depending on whether the graph is constrained to a simple graph or allowed to be a multigraph: 
\begin{align}
\mathcal{N}_{G} = 
\begin{cases}
 \displaystyle \binom{\left|\Omega_{\mathrm{in}}\right|}{M_{\mathrm{in}}} \binom{\left|\Omega_{\mathrm{out}}\right|}{M_{\mathrm{out}}} & \text{(simple graph)} \\[15pt]
\displaystyle \nHr{\left|\Omega_{\mathrm{in}}\right|}{M_{\mathrm{in}}} \nHr{\left|\Omega_{\mathrm{out}}\right|}{M_{\mathrm{out}}}  & \text{(multigraph)}
\end{cases}, \label{ORGM-NumberOfGraphs}
\end{align}
where $\left|\Omega_{\mathrm{in}}\right| = \sum_{i=1}^{N}\left( F(i) - i \right)$, $\left|\Omega_{\mathrm{out}}\right| = \binom{N}{2} - \left|\Omega_{\mathrm{in}}\right|$, and 
\begin{align}
\nHr{n}{m} \equiv \frac{(n+m-1)!}{(n-1)! m!} = \binom{n+m-1}{m}, \label{nHr-definition}
\end{align}
is the number of combinations of $m$ elements taken from $n$ elements with repetition. 
Instead of $M_{\mathrm{in}}$ and $M_{\mathrm{out}}$, the model can also be parametrized using the total number of edges $M = M_{\mathrm{in}} + M_{\mathrm{out}}$ and the density ratio defined by $\epsilon \equiv (M_{\mathrm{out}}/\left|\Omega_{\mathrm{out}}\right|)/(M_{\mathrm{in}}/\left|\Omega_{\mathrm{in}}\right|)$; the ORGM becomes a uniform model when $\epsilon=1$, while the nonzero elements are strictly confined in $\Omega_{\mathrm{in}}$ when $\epsilon=0$. 
In the ORGM, there is no finer structure within $\Omega_{\mathrm{in}}$, unlike some other order-dependent models (e.g., \cite{Grindrod2002,SongWang2014}). 
Depending on the envelope function $F(i)$, some of the vertices are statistically equivalent. 

In this paper, we focus on a simple envelope function that represents a ``banded'' structure (Fig.~\ref{fig:ORGM_adjacency_matrix}{\bf a}): 
\begin{align}
F(i) = 
\begin{cases}
i + r & (i+r \le N) \\
N & (i+r > N)
\end{cases}. \label{SimpleEnvelopFunction}
\end{align}
That is, edges are generated with a high probability within the diagonal band with ``bandwidth'' $r$ from the main diagonal. 
In this case, we have $|\Omega_{\mathrm{in}}| = r (2N-r-1)/2$. 
Note that the expected degrees of vertices are lower at both ends of the vertex sequence in this model. 
In addition, when the ORGM is constrained to simple graphs, $r$ is constrained such that $M_{\mathrm{in}} \le \left|\Omega_{\mathrm{in}}\right|$ and $M_{\mathrm{out}} \le \left|\Omega_{\mathrm{out}}\right|$ are satisfied. 
In summary, the ORGM is parametrized by $N$, $M$, $r$, and $\epsilon$.

\begin{figure*}[ht!]
  \centering
  \includegraphics[width= 1.7 \columnwidth]{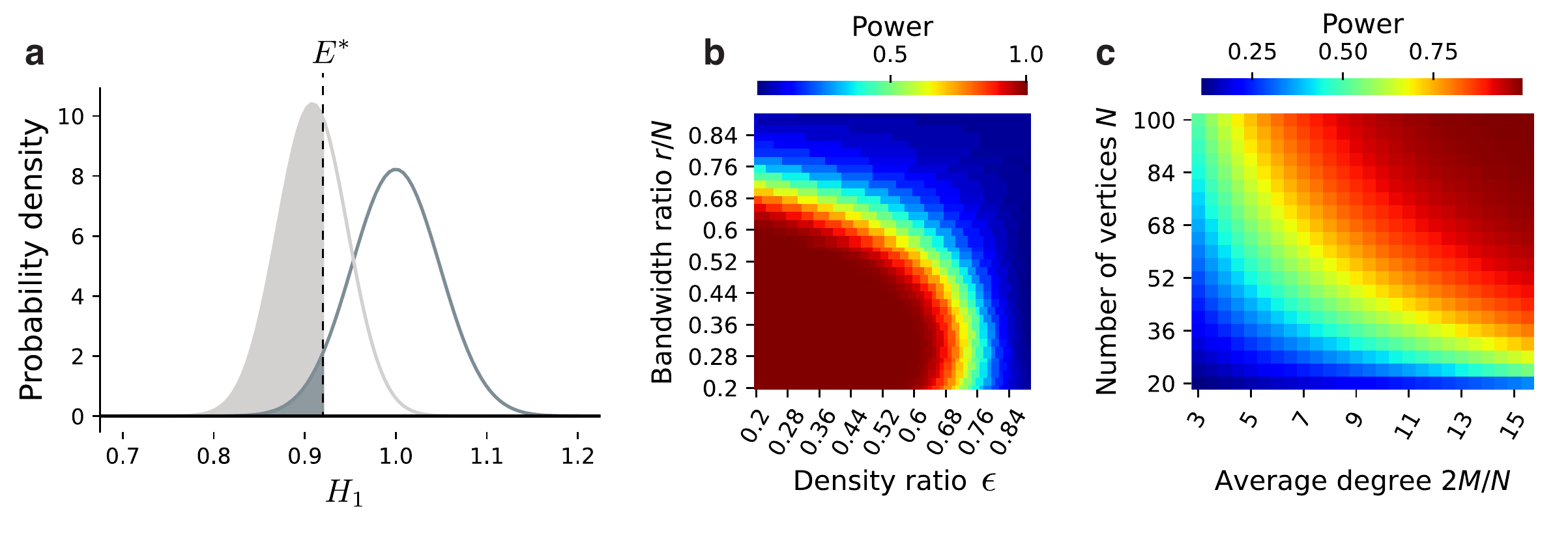}
  \caption{
  Power of the test for unoptimized sequences. 
  While the null hypothesis is the {\ER} model, the graphs are generated by the ORGM. 
  The rejection rate (dark-shaded region) of the null hypothesis and the acceptance rate (light- and dark-shaded regions) of the alternative hypothesis ($r/N = 0.75$, $\epsilon = 0$) are shown in (a) for graphs with $N=50$ and $M=200$. 
  The density plots represent the power, the true-positive rate, that we calculated analytically based on Eqs.~(\ref{PowerFunction}) and (\ref{ORGMNormalDistribution}), in (b) the $(r/N, \epsilon)$-plane ($N=50$, $M=200$) and (c) the $(N, 2M/N)$-plane ($r/N = 0.75$, $\epsilon = 0$). 
	}
  \label{fig:Power_ER_vs_ORGM}
\end{figure*}

If we set $F(i)$ such that $\Omega_{\mathrm{in}}$ constitutes a block-diagonal form, as illustrated in Fig.~\ref{fig:ORGM_adjacency_matrix}{\bf b}, the ORGM with this envelope function is equivalent to the {\SBM} with two statistically identical groups. 
Therefore, the ORGM partly overlaps with the {\SBM}, while it is flexible enough to represent a banded structure as well. 
Although Eq.~(\ref{SimpleEnvelopFunction}) is an appropriate envelope function for the purpose of this paper, the ORGM with a more complicated envelope function will be useful when it is used as an inference model. 
This is left for future work.

Using the ORGM, we investigate the extent to which the proposed test (\ref{Ztest}) is effective. 
Because it is a test of uniformity, the $p$ value of the statistical test is highly nontrivial when an adjacency matrix is close to uniformly random ($\epsilon \approx 1$ or $r/N \approx 1$ in the ORGM). 
On the other hand, there would be no need for the statistical test when we can visually confirm a sequentially local structure; the $p$ value would be trivially small.
We quantify this intuition in terms of the ORGM parameters via power analysis. 

We assume that graphs are generated from the {\ER} model as the null hypothesis, while we use the ORGM as the alternative hypothesis. 
The critical value $\stat{H}_{1} = E^{\ast}$ below which the null hypothesis is rejected is determined by 
\begin{align}
\int_{-\infty}^{\sqrt{\frac{2M (N+1)}{N-2}} (E^{\ast} - 1)} \frac{dx}{\sqrt{2\pi}} e^{-\frac{1}{2}x^{2}} = \alpha, \label{CriticalE}
\end{align}
where $\alpha$ is again the significance level; we let $\alpha=0.05$. 
As the power is the true-positive rate of the alternative hypothesis, we have 
\begin{align}
\mathrm{Power}(N,M,r,\epsilon) =  \int_{-\infty}^{E^{\ast}} dE \, \mathrm{Prob}\left[ \stat{H}_{1} = E; N, M, r, \epsilon \right], \label{PowerFunction}
\end{align}
where $\mathrm{Prob}\left[ \stat{H}_{1} = E; N, M, r, \epsilon \right]$ is the probability distribution of the $\stat{H}_{1}$ test statistic in the ORGM. 
Its specific form is derived in {\Methods}~\ref{sec:SimplegraphORGM}. 
Note that, unlike Eq.~(\ref{Ztest}), Eq.~(\ref{CriticalE}) is a one-sided test because we accept the ORGM if the null hypothesis is not true. 
In Fig.~\ref{fig:Power_ER_vs_ORGM}{\bf a}, the rejection region determined by Eq.~(\ref{CriticalE}) (dark-shaded region) and acceptance region determined by Eq.~(\ref{PowerFunction}) (light- and dark-shaded regions) are shown for a specific parameter set. 

Note that our usage of the power analysis is slightly distinct from the common usage. 
Although one usually considers the condition where a higher power can be achieved, we consider the parameter region where the power is not very high. 
For example, although the power is nearly zero when $\epsilon \approx 1$ or $r/N \approx 1$, it is certainly the case where we wish to try the statistical test; note that we do not know the model parameters in practice, and we cannot figure it out from the visual inspection. 
Another distinction from the common usage is that we are not certain about the alternative hypothesis. 
Although we consider the ORGM as the alternative hypothesis, this is only one of many models that can generate a sequentially local structure. 
Therefore, even in the parameter region where the power is low, it does not imply the test is useless there. 
In such a region, we should execute the test and confirm whether the locality is significant or not. 

Figure \ref{fig:Power_ER_vs_ORGM}{\bf b} shows the $r$--$\epsilon$ dependency of the power for a given set of $N$ and $M$. 
Furthermore, Fig.~\ref{fig:Power_ER_vs_ORGM}{\bf c} shows the $N$--$M$ dependency of the power when $r$ and $\epsilon$ are fixed. 
As observed in Fig.~\ref{fig:Power_ER_vs_ORGM}{\bf b}, the power is not nearly 1 for $r/N \gtrsim 0.75$ even when $\epsilon \simeq 0$.
This result implies that the present test is expected to be meaningful (i.e., the $p$ value would not be extremely small) when a graph has edges between vertices with sequential distance $|\pi_{i} - \pi_{j}| \gtrsim 0.75 N$. 
Similarly, Fig.~\ref{fig:Power_ER_vs_ORGM}{\bf c} indicates that the present test is expected to be meaningful for small sparse graphs. 
The plot represents the quantitative relationship between the sparsity and graph size. 
Figure \ref{fig:Power_ER_vs_ORGM}{\bf b} also indicates the finite-size detectability of the ORGM. 
For example, as long as the density ratio $\epsilon$ is sufficiently large (e.g., $\epsilon \gtrsim 0.8$ in Fig.~\ref{fig:Power_ER_vs_ORGM}{\bf b}), the null hypothesis of the {\ER} model is rarely rejected, that is, the power is close to zero, for any bandwidth $r$.

\section{Sequential locality of optimized sequences \label{sec:OptimizedTest}}
\begin{figure}[t]
  \centering
  \includegraphics[width= 0.99 \columnwidth]{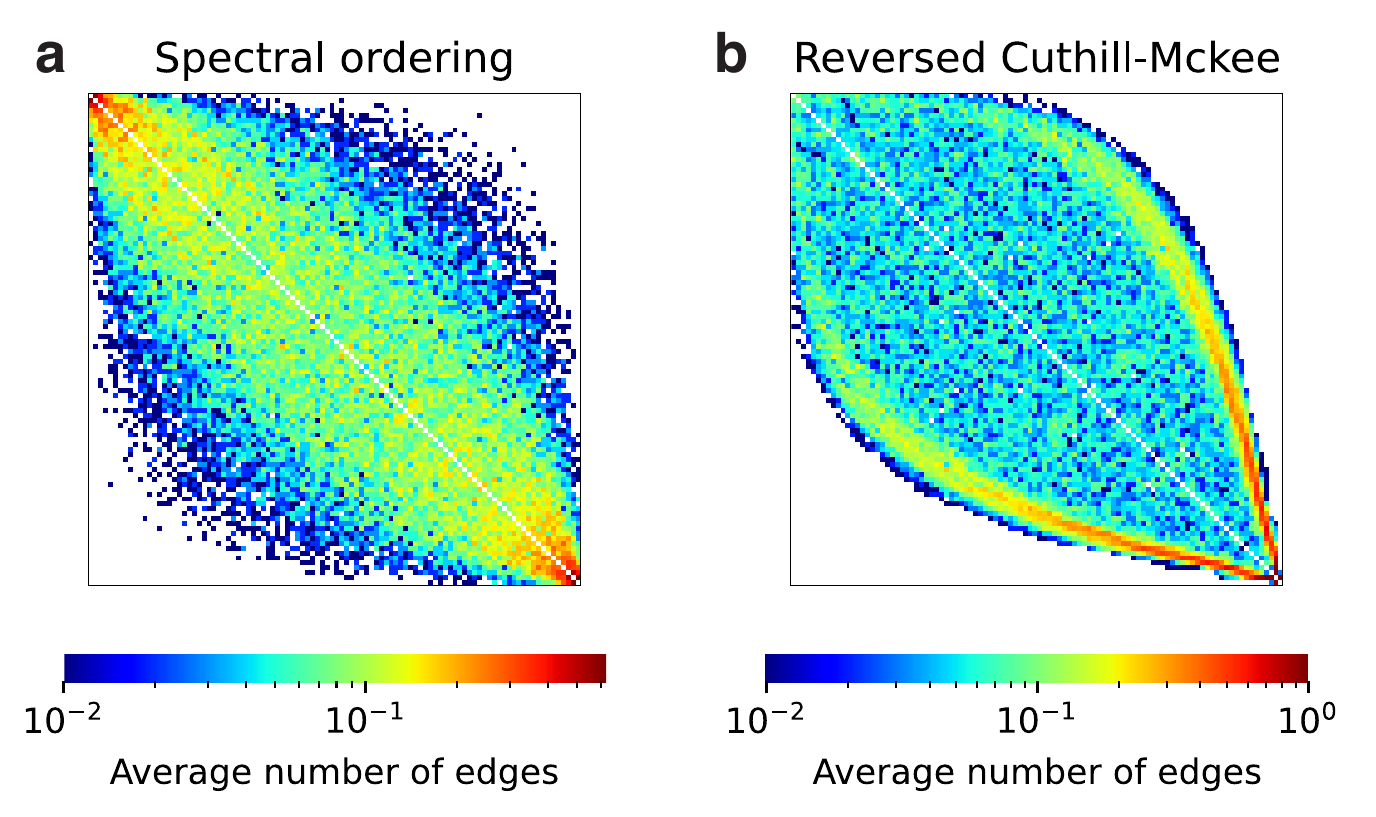}
  \caption{
  Average of the adjacency matrices for the {\ER} model in optimized vertex orderings. 
  The vertex sequences are optimized using (a) the spectral ordering and (b) reversed Cuthill-McKee algorithm. 
  In each case, we set $N=100$ and $M=250$. 
  Each cell in a density plot represents the average value of the adjacency matrix element in $100$ samples. 
	}
  \label{fig:Optimized_ERgraphs}
\end{figure}

The statistical test in the previous section assumed that the vertex sequence is not optimized. 
This assumption is important because the results in the previous section imply that the test is not suitable for cases with optimized vertex sequences. 
For example, for the \textit{political books} dataset with the adjacency matrix in Fig.~\ref{fig:Polbooks}{\bf c}, testing based on Eq.~(\ref{H1NullDistributionExact}) is not a fair comparison because the vertex sequence is optimized so that the adjacency matrix exhibits strong sequential locality. 
In fact, even graphs generated by the {\ER} model can exhibit significant sequential locality under Eq.~(\ref{H1NullDistributionExact}) when the vertex sequence is optimized. 
In Fig.~\ref{fig:Optimized_ERgraphs}, we show the average adjacency matrices of the {\ER} model in which the vertices are ordered using different algorithms: the spectral ordering \cite{DingHe2004} (Fig.~\ref{fig:Optimized_ERgraphs}{\bf a}) and the reversed Cuthill-McKee algorithm \cite{golub1996matrix,Higham2003} (Fig.~\ref{fig:Optimized_ERgraphs}{\bf b}).

A simple approach to dealing with this problem is to calculate an empirical distribution of the test statistic for a set of optimized vertex sequences obtained from randomized graphs, e.g., graphs obtained via rewiring of an input graph. 
If the observed test statistic of the actual graph is sufficiently far apart from those of the randomized samples, we conclude that the graph has a sequentially local structure that cannot be commonly obtained by randomized counterparts. 
However, a disadvantage of this simple approach is that, in many cases, it can only be used for descriptive purposes. 
Note that the empirical distribution of the test statistics is generally affected not only by the intrinsic structure of the graph, but also by the optimization algorithm used to obtain the vertex sequence. 
In other words, the null hypothesis is often highly complicated. 
Therefore, even if we conduct a formal statistical test, we can hardly interpret what the resulting $p$ value really implies (see {\Methods}~\ref{sec:NullHypothesisFallacy} for more discussion). 
Although the null hypothesis can be algorithm-independent when an optimization is executed exactly, such a test would be computationally infeasible.

Herein, instead of considering a random model for a matrix before optimization and including algorithmic dependencies in the null hypothesis, we consider a random model for a matrix after optimization and exclude algorithmic dependencies from the null hypothesis. 
To this end, we propose a statistical test in which the ORGM is used to define a null hypothesis. 
We let $N$ and $M$ be equal to the observed numbers of vertices and edges in the dataset, respectively, with which we obtain the maximum likelihood estimators (MLEs) for the bandwidth $r$ and the density ratio $\epsilon$ (we describe the details of the MLEs in {\Methods}~\ref{sec:MLE_r}. 
Then, we assess whether the elements of the optimized adjacency matrix are uniformly distributed within $\Omega_{\mathrm{in}}$. 
In other words, we ignore the observed elements in $\Omega_{\mathrm{out}}$ and compute the $\stat{H}_{1}$ value (i.e., ``in-envelope $H_1$"). 
Using the theoretical mean $\mathbb{E}_{\mat{A}}[\stat{H}_1(\mathbf{\pi}; \mat{A})]$ (Eq.~(\ref{ORGM-1stMoment})) and second moment $\mathbb{E}_{\mat{A}}[\stat{H}^{2}_1(\mathbf{\pi}; \mat{A})]$ (Eq.~(\ref{ORGM-2ndMoment})) of the $\stat{H}_{1}$ statistic, we can analytically compute the $p$ value of the observed graph. 
Importantly, in contrast to the aforementioned simple approach, there is no algorithmic uncertainty at the stage of the statistical test because the null hypothesis is specified exclusively by the fitted ORGM without any influence from the optimization algorithm. 

This test can be viewed as a variant of the test for unoptimized sequences in Sec.~\ref{sec:UnoptimizedTest}. 
Instead of testing a uniform structure in the entire matrix space, we execute it in a subspace of the adjacency matrix. 
However, the implication of the $p$ value is very different from that in Sec.~\ref{sec:UnoptimizedTest}. 
Here, a smaller $p$ value implies that the graph is expected to have a finer local structure than that assumed in the ORGM. 

\begin{table*}[tbh]
\caption{Description of datasets. The $p$ value for the two-sided test is obtained from the ORGM hypothesis based on $\Omega_{\rm in}$ (ORGM), and the $z_1$ factor for the random sequence hypothesis is defined as $z_1/\sqrt{{\rm{Var}}_{\ket{\pi}}[z_1({\ket{\pi}};A)]}$ (rand. seq.).}
\begin{center}
\begin{tabularx}{\textwidth}{lcccccXc}
 \hline
   \multirow{2}{*}{Dataset} &   \multirow{2}{*}{$N$} &   \multirow{2}{*}{$M$} & \multirow{2}{*}{$r^{\ast}$} & \multirow{2}{*}{\parbox{1.5cm}{$p$ value  (ORGM)}}  &  \multirow{2}{*}{\parbox{1.8cm}{ $z_1$ factor  (rand.\ seq.) }} &\multirow{2}{*}{Data description} & \multirow{2}{*}{Refs.} \\
   & & & & & &  & \\
   \hline 
          Tribes &   16 &    58 &           8 &  0.126   & $-3.563$  & Friendship network of tribes in New Guinea&  \cite{read1954cultures}     \\           
          Montreal &   29 &    75 &           8 & $<$0.001  & $-1.275$&   Relationships between gangs, obtained from the Montreal police department's central intelligence database &  \cite{descormiers2011alliances} \\            
          States &   49 &   107 &           6 &  $<$0.001  & $-9.017$  &  Network of contiguous states in the United States &       \cite{knuth1993stanford} \\                  
          Highschool &   70 &   366 &           9 & $<$0.001 & $-10.884$ & Friendship network of male students in a high school in Illinois. & \cite{coleman1964introduction} \\                      
          Polbooks &  105 &   441 &          21 & $<$0.001  & $-13.551$  &   Copurchase network of books about US politics. &       \cite{polbook} \\                       
          Adjnoun &  112 &   425 &          30 & $<$0.001   &  $-3.323$  & Word adjacencies of common adjectives and nouns in the novel \textit{David Copperfield} &  \cite{newman2006finding} \\             
          Football &  115 &   613 &          23 &  $<$0.001  &   $-3.476$ &  Network of American football games between Division IA colleges. &   \cite{girvan2002community}, \cite{evans2010clique} \\  
          Ugandan &  181 &   774 &          70& $0.968$  &  $-3.888$  & Social network in a Ugandan village  &        \cite{chami2017social} \\               
          Celegans &  297 &  2359 &          82 & $<$0.001  &  $-12.655$ & Neural connections of the \textit{C.\ elegans} nematode &       \cite{white1986structure} \\  
           Transport &  369 &   441 &          15 & $<$0.001  &  $-10.846$ & Network of London train stations: Underground, Overground and DLR & \cite{de2014navigability} \\
           \hline 
 \end{tabularx}
 \end{center}
 \label{tab:dataset}
\end{table*}

\begin{figure*}[t!]
  \centering
   \includegraphics[width= 1.6 \columnwidth]{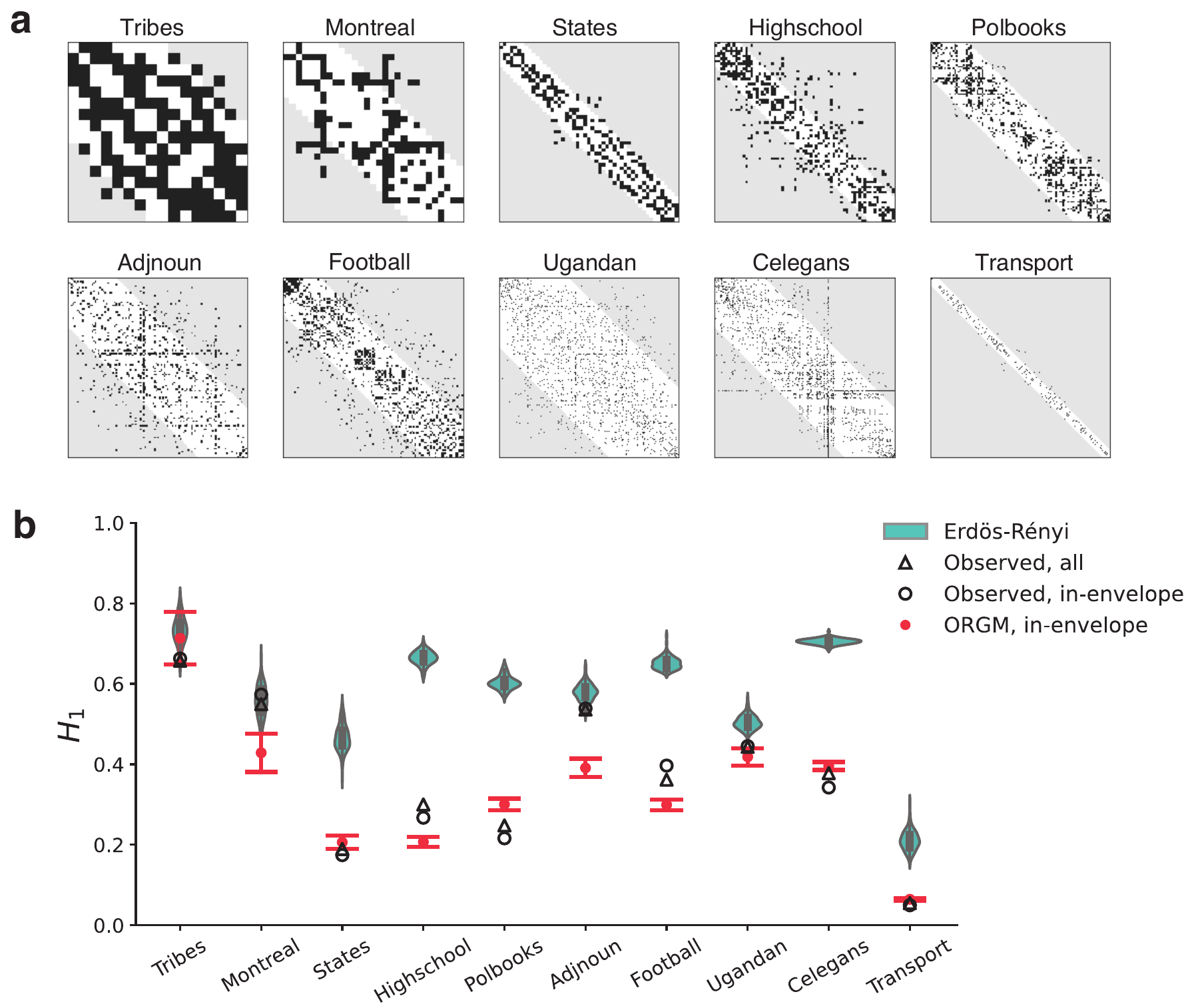}
 \caption{Statistical tests for optimized sequence using real-world datasets. (a) Optimized adjacency matrices. The white and gray areas denote $\Omega_{\mathrm{in}}$ and $\Omega_{\mathrm{out}}$, respectively, obtained via the estimated bandwidth $r^{\ast}$. 
 The black cells represent the elements with $A_{ij}=1$. 
  (b) $\stat{H}_{1}$ values for different datasets. 
  For each data, the triangle represents the observed $\stat{H}_{1}$ value (``Observed, all"), while the violin plot represents the kernel density estimate \cite{seaborn-violinplot} of the empirical $\stat{H}_{1}$ distribution based on 100 optimized samples of the {\ER} model (``Erd\H{o}s-R\'{e}nyi"). 
  The circles and error-bar plots describe the statistical tests using the ORGM. 
  Each circle is the observed $\stat{H}_{1}$ value within $\Omega_{\mathrm{in}}$ (``Observed, in-envelope"). 
  The corresponding error bar plot (red) represents the mean and $95\%$ confidence interval of the fitted ORGM in which the elements in $\Omega_{\mathrm{out}}$ are ignored (``ORGM, in-envelope").} 
  \label{fig:multidata_optimized_test_adjacency}
\end{figure*}

\begin{figure*}[t!]
  \centering
    \includegraphics[width= 1.99\columnwidth]{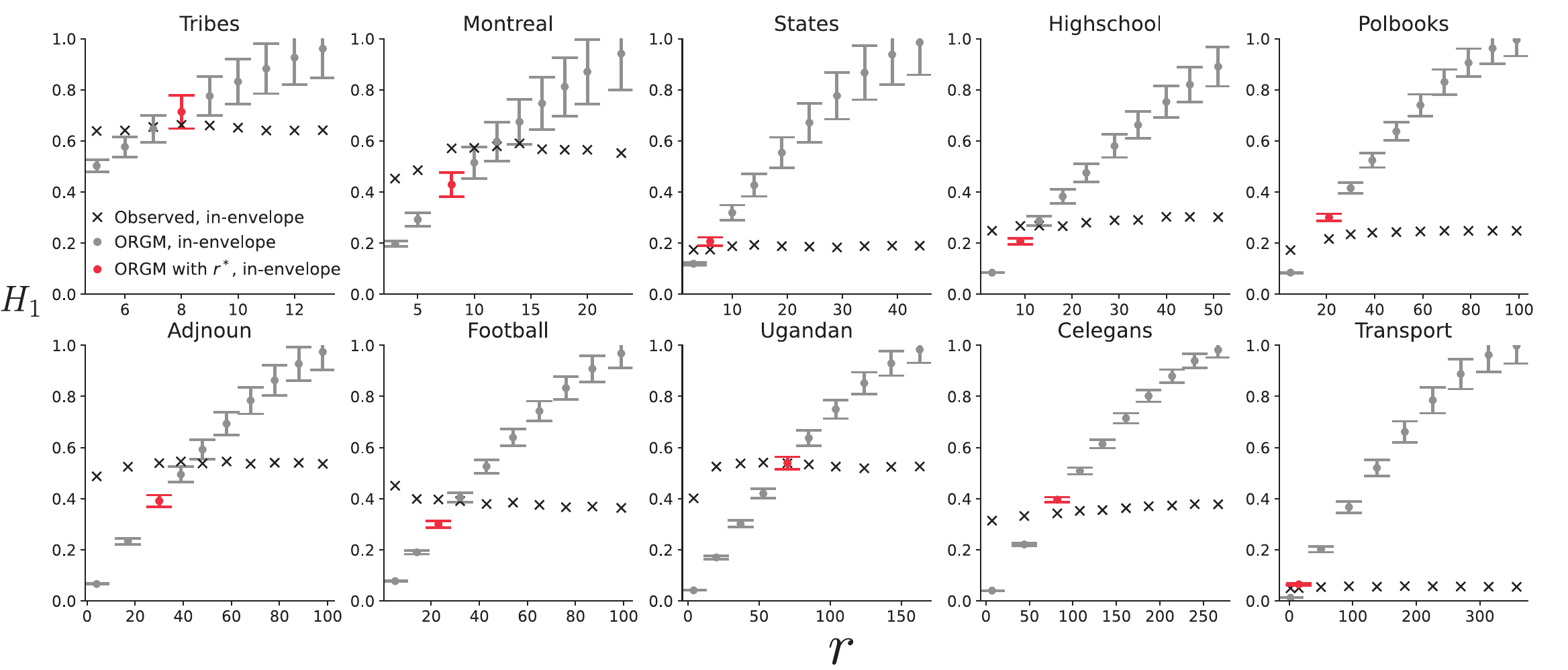}
 \caption{Confidence intervals of $\stat{H}_{1}$ based on the ORGM for different values of the bandwidth $r$. 
 Each confidence interval is obtained using the matrix elements in $\Omega_{\rm in}$ for given $r$, i.e., the interval for the in-envelope estimate. 
 The red error bar represents the confidence interval with $r^{\ast}$, the value employed in Fig.~\ref{fig:multidata_optimized_test_adjacency}{\bf{b}}. 
 The crosses represent the observed in-envelope $\stat{H}_{1}$ values.
 These plots represent the sensitivity of the statistical test with respect to the bandwidth estimate. }    
   \label{fig:r-sweep}
\end{figure*}

The (masked) adjacency matrices for different real-world datasets are shown in Fig.~\ref{fig:multidata_optimized_test_adjacency}{\bf a}, for which we use the estimated bandwidth $r^{\ast}$, and the corresponding test statistics are plotted in Fig.~\ref{fig:multidata_optimized_test_adjacency}{\bf b} (see Table~\ref{tab:dataset} for a description of the datasets).  
All the datasets in Table~\ref{tab:dataset} are downloaded from the network repository Netzschleuder~\cite{peixoto2020netzschleuder}. 
The statistical tests based on $\Omega_{\mathrm{in}}$ allow us to classify the datasets into four types. 
\begin{description}
\item[ I ] The observed in-envelope $\stat{H}_{1}$ is lower than the confidence interval (CI) of the ORGM hypothesis (see Table~\ref{tab:dataset} for the corresponding $p$ value), meaning that the edges in $\Omega_{\rm in}$ exhibit a finer structure in the sense of sequential locality (e.g., States, Polbooks, Celegans, and Transport). 
\item[ II ] The observed in-envelope $\stat{H}_{1}$ is within the CI, meaning that we cannot reject the hypothesis that the edges in $\Omega_{\rm in}$ are connected uniformly at random as suggested by the ORGM; i.e., the dataset is well-characterized solely by the sequentially-local structure (e.g., Tribes and Ugandan).
\item[ III ] The observed in-envelope $\stat{H}_{1}$ is above the CI, yet the observed $\stat{H}_{1}$ based on all the edges is far below those of the {\ER} model, indicating that the dataset is not well characterized by the ORGM. However, the graph has some extent of sequential locality compared with its randomized counterpart (e.g., Highschool and Football). 
\item[ IV ] The observed $\stat{H}_{1}$ value for the entire adjacency matrix can be typically achieved by the {\ER} model, and thus the graph does not exhibit a strong sequential locality compared with the randomized counterpart (e.g., Montreal and Adjnoun).
\end{description}
The number of edges in $\Omega_{\rm in}$ can vary as $r$ changes, and therefore, the in-envelope CI may also change accordingly. 
When the evaluated $p$ value is unfairly low owing to the fact that $r^{\ast}$ is too large (which can be visually confirmed from the adjacency matrix as long as the dataset is not too large), one ought to confirm whether the in-envelope CI is highly sensitive to $r$ (Fig.~\ref{fig:r-sweep}). 
Note that the MLE $r^{\ast}$ is provided only to determine a plausible bandwidth systematically and automatically. 
It is ultimately the analyst's choice as to which region to consider as $\Omega_{\rm in}$. 

The present statistical test implies a connection between envelope reduction and community detection. 
As mentioned in Sec.~\ref{sec:Introduction}, a graph can have a community structure either with or without a banded structure. 
Our ORGM hypothesis will be rejected in both of these cases owing to the heterogeneity of edge density characterized by the community structure (these cases typically fall within either type I or type III under the above criteria). 
Therefore, it is natural that datasets often used for a benchmark test in community detection (e.g., Polbooks, Celegans, and Football) exhibit extremely low $p$ values.

For a more quantitative insight, let us consider the upper bound of the in-envelope $\stat{H}_{1}$ value. 
The $\stat{H}_{1}$ statistic can be regarded as a rescaled average sequential distance between connected vertices because we have
\begin{align}
\frac{1}{M}\sum_{i<j} A_{ij} \left|\pi_{i} - \pi_{j}\right|
= \frac{N+1}{3} \stat{H}_{1}(\mat{A}, \ket{\pi}). 
\end{align}
As the average sequential distance within the estimated envelope cannot be larger than $r^{\ast}$, we have $\max \stat{H}_{1} = 3 r^{\ast}/(N+1)$ as an upper bound; only when every connected vertex pair is separated by $r^{\ast}$, does it actually become the maximum.
Using Eq.~(\ref{ORGM-1stMoment}) in {\Methods}~\ref{sec:SimplegraphORGM}, we have 
\begin{align}
& \frac{\max \stat{H}_1}{\mathbb{E}_{\mat{A}}\left[\stat{H}_1\right]} 
= \frac{3r (2N-r-1)}{(r+1)(3N-2r-1)}. \label{MaxAverageRatio}
\end{align}
When $N \gg 1$, this fraction is a monotonically increasing function with respect to $r^{\ast}/N$ ($0 < r^{\ast}/N < 1$), indicating that $\max \stat{H}_1$ becomes relatively larger than $\mathbb{E}_{\mat{A}}\left[\stat{H}_1\right]$ when the bandwidth ratio $r^{\ast}/N$ is large. 
Using the max-average ratio (\ref{MaxAverageRatio}), in addition to the assessment of statistical significance, we can evaluate how close the observed $\stat{H}_1$ value is to its upper bound. 
The variance $\mathrm{Var}_{\mat{A}}\left[ \stat{H}_{1}(\mat{A}, \ket{\pi}) \right]$, however, is not solely described by $r^{\ast}/N$.

\section{Sequential locality of random vertex sequences \label{sec:RandomSequence}}
So far, we have considered tests in which a vertex sequence $\ket{\pi}$ is given and assessed the statistical significance of graphs, or adjacency matrices. 
Here we assess whether the inferred sequence is a significantly better choice than a random guess given a graph in the sense of stronger sequential locality. 
To this end, we consider random sequences in which every possible sequence occurs with equal probability. 

The mean and variance of the $z_1$ test statistic for random sequences are given by 
\begin{align}
&\mathbb{E}_{\ket{\pi}}\left[ z_{1}(\mat{A}, \ket{\pi}) \right] = 0, \label{MeanRandomSequenceH1}\\
& \mathrm{Var}_{\ket{\pi}}\left[ z_{1}(\mat{A}, \ket{\pi}) \right] \notag\\
&= \frac{N+1}{N-2}  \left(
\frac{5N-8}{5(N+1)} 
+ \frac{M_{3} (N-4)}{5M (N+1)}
- \frac{2M}{5 (N+1)}
\right), 
\label{VarianceRandomSequenceH1}
\end{align}
where $M_{3}$ is the total number of connected edge pairs, or wedges. 
The detailed derivations of Eqs.~(\ref{MeanRandomSequenceH1}) and (\ref{VarianceRandomSequenceH1}) can be found in {\Methods}~\ref{sec:RandomSequence-Appendix}. 
Importantly, the mean value is zero for an arbitrary graph, and the variances depend only on the set of macroscopic quantities $(N, M, M_{3})$. 
In other words, it has no dependency on microscopic quantities, such as the degree sequence. 

\begin{figure*}[t!]
  \centering
  \includegraphics[width= 1.99 \columnwidth]{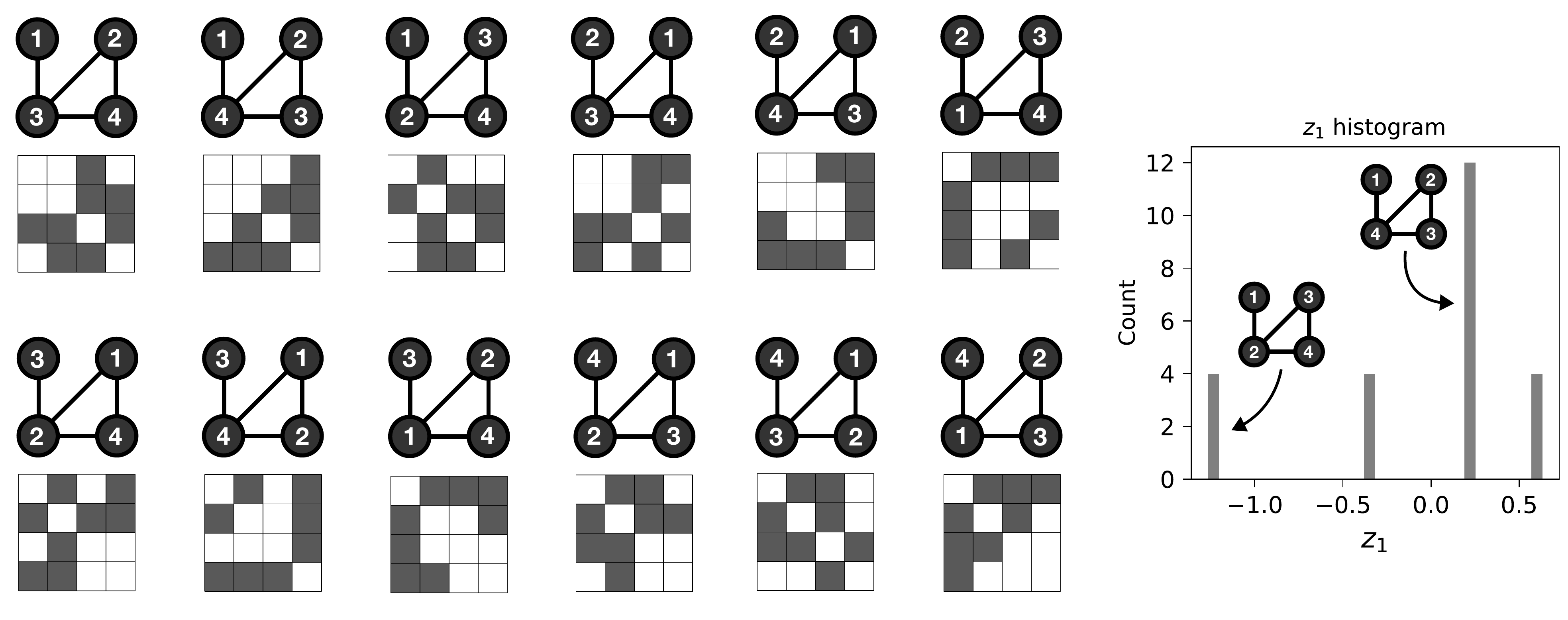}
  \caption{
  List of all distinguishable isomorphisms and adjacency matrices (white cells for $A_{ij}=0$ and black cells for $A_{ij}=1$) of a small graph and the histogram of the $z_{1}$ test statistic with respect to all vertex sequences. 
  Among $4! = 24$ permutations of the vertex sequence, $12$ of them are distinct isomorphisms. Therefore, every graph is counted twice ($|\mathrm{Aut}(G)|=2$) in the histogram, although it has no effect on the assessment of statistical significance.
	}
  \label{fig:test_sequences}
\end{figure*}

Unfortunately, deriving an analytical form of the probability distribution for the $z_1$ test statistic is not trivial. 
It is expected from the calculation of the second moment that the higher-order moments depend on the total number of triangles and other types of motifs. 
Therefore, it is not straightforward to compute the $p$ value and conduct a statistical test because we cannot generally assume that the test-statistic distribution is nearly normal. 
However, unlike the test of adjacency matrices with optimized vertex sequences, there is no fundamental difficulties. 
We can obtain the exact distribution by computing $z_1$ for all possible vertex sequences as long as it is computationally feasible (see Fig.~\ref{fig:test_sequences} for a simple example). 
If not, we can estimate the distribution via uniform sampling of vertex sequences. 

We emphasize that this hypothesis testing assesses the quality of vertex sequences for a given graph but does not test whether a graph is sequentially local or not. 
For example, even when a graph is generated from a uniform random graph model, the $p$ value for an optimal sequence is exactly zero by definition. 
In the example of Fig.~\ref{fig:test_sequences}, the sequences with $z_{1}=-1.12$ have a $p$ value equal to zero. 
Empirically, unless an optimization algorithm works very poorly, the hypothesis of a random sequence is often rejected when a well-permuted vertex sequence is tested. 
Therefore, similar to the test based on the {\ER} model in Eq.~(\ref{H1NullDistributionNormal}), the null hypothesis is more suitable for testing unoptimized vertex sequences than for testing optimized sequences. 

In Table~\ref{tab:dataset}, we show the results for real-world datasets in which the vertex sequences are not optimized. 
Instead of the $p$ value, we show the factor $z_{1}/\sqrt{\mathrm{Var}_{\ket{\pi}}\left[z_{1}(\mat{A}, \ket{\pi}) \right]}$, which we refer to as the $z_{1}$ factor, for each dataset to indicate the extent to which the test statistic under the original vertex ordering is different from the typical scale of random sequences.
It is observed that the original vertex sequences in most of these datasets are not likely to be sampled uniformly randomly. 

\begin{figure}
  \centering
  \includegraphics[width= 0.99\columnwidth]{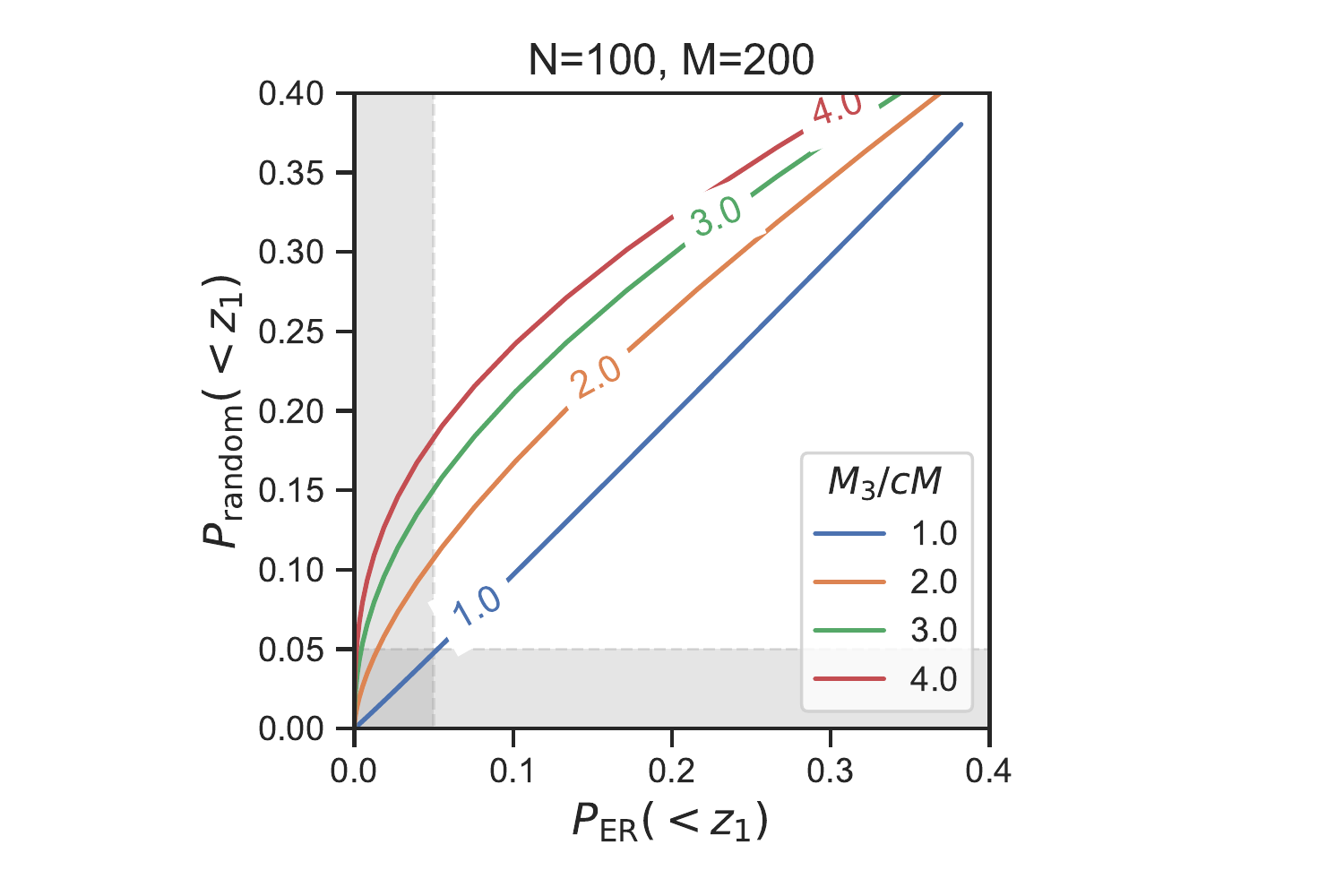}
  \caption{
Parametric plot for the assessment of $z_{1}$ statistics under the {\ER} hypothesis and under random sequences.
 $P_{\mathrm{ER}}(< z_{1})$ and $P_{\mathrm{random}}(< z_{1})$ denote the cumulative probabilities for the {\ER} and random sequence hypotheses, respectively.
Each colored line is obtained by sweeping the value of $z_{1}$ for a given $M_{3}/(cM)$.
The boundary of the vertical (horizontal) shaded area represents the points at which the cumulative probability $P_{\mathrm{ER}}(< z_{1})$ ($P_{\mathrm{random}}(< z_{1})$) is $0.05$.  
We set $N=100$ and $M=200$ ($c = 2M/N$).  
	}
  \label{fig:ERvsRandomSeqComparison}
\end{figure}

It is worth mentioning the difference between the test-statistic distribution for random sequences and that under the {\ER} hypothesis (Eq.~(\ref{H1NullDistributionNormal})). 
Note that although the mean value is always zero in both cases, the variance (\ref{VarianceRandomSequenceH1}) may not be equal to unity. 
In fact, we have $M_{3}/M \gg M/N$ unless most of the vertices have very low degrees, and thus, the variance (\ref{VarianceRandomSequenceH1}) is typically considerably larger than unity.
A consequence of having a variance larger than unity is that a graph associated with a randomized vertex sequence is likely to be identified as having significant sequential locality when the adjacency matrix is assessed based on the {\ER} model. 
The parametric plot in Fig.~\ref{fig:ERvsRandomSeqComparison} quantitatively shows this tendency. As we make $z_{1}$ smaller, both of the cumulative probabilities $P_{\mathrm{ER}}(< z_{1})$ and $P_{\mathrm{random}}(< z_{1})$ decrease. However, $P_{\mathrm{ER}}(< z_{1})$ decreases more rapidly. Therefore, whereas the adjacency matrix with a small value of $z_{1}$ commonly emerges within the graphs with randomized vertex sequences (because $P_{\mathrm{random}}(< z_{1})$ is relatively large), it can be assessed as statistically significant in the test under the {\ER} model (because $P_{\mathrm{ER}}(< z_{1})$ is relatively small). 
This tendency explains why the instances of the {\ER} model can exhibit a strong sequential locality if one carefully chooses the vertex sequence, as shown in Fig.~\ref{fig:Optimized_ERgraphs}.

Before concluding this section, let us mention the overcounting of automorphisms in random sequences. 
The ordering of each vertex sequence is regarded as a permutation of the original ordering. 
Among all possible permutations, the subgroup that yields the same adjacency matrix as the original matrix constitutes the automorphism group \cite{beineke2004topics}, that is, 
\begin{align}
\{ \ket{\pi} | \mat{\Perm}_{\pi}^{\top} \mat{A} \mat{\Perm}_{\pi} = \mat{A} \} =: \mathrm{Aut}(G),
\end{align}
where the vertex set $V$ of graph $G$ is indexed by the raw indices $I$. 
If we assume that distinct adjacency matrices are drawn uniformly randomly, the sequences yielding the identical adjacency matrix are overcounted in Eqs.~(\ref{MeanRandomSequenceH1}) and (\ref{VarianceRandomSequenceH1}). 
However, as described below, the number of overcounts is equal for every distinct adjacency matrix. 
This implies that overcounting within each automorphism group has no effect on the probability distribution after all (see Fig.~\ref{fig:test_sequences} for an example). 

To evaluate the number of overcounts, we use the Lagrange theorem \cite{judson2009abstract}.
Let us consider a permutation $\ket{\tau}$ such that $\mat{\Perm}_{\tau}^{\top} \mat{A} \mat{\Perm}_{\tau} =: \mat{A}^{\prime} \ne \mat{A}$ and we denote the graph $G^{\prime} = (V^{\prime}, E^{\prime})$ in which the vertex set $V^{\prime}$ is indexed  based on the permutation $\ket{\tau}$. 
The coset of the automorphism group $\mathrm{Aut}(G)$ with $\ket{\tau}$ reads 
\begin{align}
\ket{\tau} \cdot \mathrm{Aut}(G) 
&= \{ \ket{\tau} \cdot \ket{\pi} | \mat{\Perm}_{\tau}^{\top} \mat{\Perm}_{\pi}^{\top} \mat{A} \mat{\Perm}_{\pi} \mat{\Perm}_{\tau} = \mat{A}^{\prime} \} \notag\\
&= \{ \ket{\tau} \cdot \ket{\pi} | \left( \mat{\Perm}_{\tau}^{\top} \mat{\Perm}_{\pi} \mat{\Perm}_{\tau} \right)^{\top} \mat{A}^{\prime} \mat{\Perm}_{\tau}^{\top} \mat{\Perm}_{\pi} \mat{\Perm}_{\tau} = \mat{A}^{\prime} \} \notag\\
&= \mathrm{Aut}(G^{\prime}). 
\end{align}
Hence, a coset of $\mathrm{Aut}(G)$ constitutes the automorphism group $\mathrm{Aut}(G^{\prime})$ with respect to $G^{\prime}$. 
The Lagrange theorem states that the cardinality of every coset of a subgroup ($\mathrm{Aut}(G)$) is equal to the the cardinality of the subgroup, indicating that the number of permutations yielding the the identical adjacency matrix is equal for every distinct adjacency matrix.

\section{Discussion \label{sec:Discussion}}
Matrix optimization problems with respect to row- and column-permutations have been extensively studied in the literature.
The main contribution of our study is that we developed a hypothesis testing framework to assess the sequentially local structure of graphs along a specified vertex sequence. 
Essentially, our work provides a statistical foundation for envelope reduction (or minimum linear arrangement), which has been formulated as an optimization problem, just as the stochastic block modeling placed the minimum-cut and related problems in a framework of statistical inference~\cite{BickelChen2009,Bianconi2009,Lei2016,Young2017,Peel2017,AbbleReview2017,Peixoto2017tutorial}. 

The statistical test for sequential locality can be useful even when no optimization algorithms are applied. 
In constructing empirical graph data, vertex indices often reflect an intrinsic vertex ordering such as chronological ordering unless they are carefully labeled to be random. 
Even when no evident characteristic structure is identified through a visual inspection of the adjacency matrix, it is still a nontrivial task to investigate whether such an ordering is statistically deemed to be random or exhibits a significant sequential locality. 
The proposed test and statistical assessment for unoptimized vertex sequences are useful tools in such cases. 
Looking at this differently, we would not learn much from the test we proposed if a vertex sequence clearly achieves a strong sequential locality. 
We quantified when and to what extent the proposed test is effective via power analysis in Sec.~\ref{sec:ORGM}. 

When the vertex sequence is optimized, the statistical test for unoptimized vertex sequences is not directly applicable because a strong sequential locality can be achieved even when a graph is generated uniformly randomly (Sec.~\ref{sec:OptimizedTest}). 
To this end, we used the ORGM that divides the space of an adjacency matrix into the region $\Omega_{\mathrm{out}}$ where the elements are typically zero and the region $\Omega_{\mathrm{in}}$ where the same test as for the unoptimized test is applicable. 
Although the ORGM is a special case of more general models such as exponential random graph models \cite{Robins2007,Lusher2013} or latent space models \cite{Hoff2002,Handcock2007}, it is more tractable because its statistics can be calculated in a combinatorial manner. 
The test for optimized vertex sequences can be useful even when we do not perform optimization by ourselves. 
For example, the original vertex indices in a dataset may already be non-random, if not optimal. 

We emphasize that our statistical tests have explicit dependencies on the total number of vertices $N$ and edges $M$. 
Moreover, $N$ and $M$ are strictly constrained in the null models to make them consistent with the observed data. 
Although we used normal approximations based on the central limit theorem, the estimated distribution is moderately accurate unless the dataset is extremely small and/or dense. 
In large-scale graphs, on the other hand, typical graph instances generated from the {\ER} model cover only a small fraction of the entire space of graph instances. 
Consequently, the $p$ value for the test of sequential locality can easily be small when $N$ is large, that is, the power of a test tends to be very high. 
This phenomenon simply indicates that the finite-size effect of graphs is considered in the statistical assessment. 

As mentioned in Sec.~\ref{sec:ORGM}, the vertices at both ends of the vertex sequence have relatively low degrees in the ORGM. This tendency could be eliminated by imposing a periodic boundary condition in the sequence. However, the boundary effect can be important for the assessment of adjacency matrices because adjacency matrices do have boundaries.

Although we focused only on the affinity matrix $\mat{J}$ with the sequential distance, we also briefly investigated the performance of the test statistic using the logarithmic semimetric, which we refer to as $\stat{H}_{G}$, in {\Methods}~\ref{sec:HGstatistic}. 
We confirmed that the test with $\stat{H}_{G}$ has a higher power than that of $\stat{H}_{1}$. 
However, we conclude that $\stat{H}_{1}$ is more useful because its analytical estimates of moments are more tractable than those of $\stat{H}_{G}$. 

Graphs with high-degree vertices tend to have significant sequential locality in our tests for graphs, particularly in the test with unoptimized vertex sequences.
Because a high-degree vertex is connected to relatively distant vertices, they tend to increase the non-local nature of the graph.
When we wish to eliminate such an effect due to degree distributions, we should consider a degree-corrected random graph model as a null model. 
This is left for future work. 

The code for the statistical tests is available on Github \cite{GithubURL}.

\appendix

\section{Moments and standardization of the $\stat{H}_{1}$ statistic}\label{sec:ERrandom-H1-Appendix}
The first and second moments of the random variable $\mathsf{X}$ that obey the triangular distribution are 
\begin{align}
&\mathbb{E}\left[ \mathsf{X} \right] = \sum_{k=1}^{N} k \, \frac{2(N - k)}{N(N-1)} = \frac{N+1}{3}, \label{ERH1_1stMoment-Appendix} \\
&\mathbb{E}\left[ \mathsf{X}^{2} \right] = \sum_{k=1}^{N} k^{2} \, \frac{2(N - k)}{N(N-1)} = \frac{N(N+1)}{6}. \label{ERH1_2ndMoment-Appendix}
\end{align}
According to the central limit theorem, the following quantity follows the standard normal distribution: 
\begin{align}
&\sqrt{\frac{M}{\mathrm{Var}\left[ \mathsf{X} \right]}} \left( \frac{1}{M}\sum_{m=1}^{M} \mathsf{X}_{m} - \mathbb{E}\left[ \mathsf{X} \right] \right) \notag\\
&= \mathbb{E}\left[ \mathsf{X} \right] \sqrt{\frac{M}{\mathrm{Var}\left[ \mathsf{X} \right]}} \left( \mathsf{H}_{1} - 1 \right) 
= \sqrt{\frac{2M (N+1)}{N-2}} \left( \mathsf{H}_{1} - 1 \right). \label{H1Standardization-Appendix}
\end{align}

\section{Comparison of the $\stat{H}_{1}$ and $\mathsf{H}_{1}$ statistics with a specific example}\label{sec:XmStatistic}
\begin{figure*}[t]
  \centering
  \includegraphics[width= 1.99 \columnwidth]{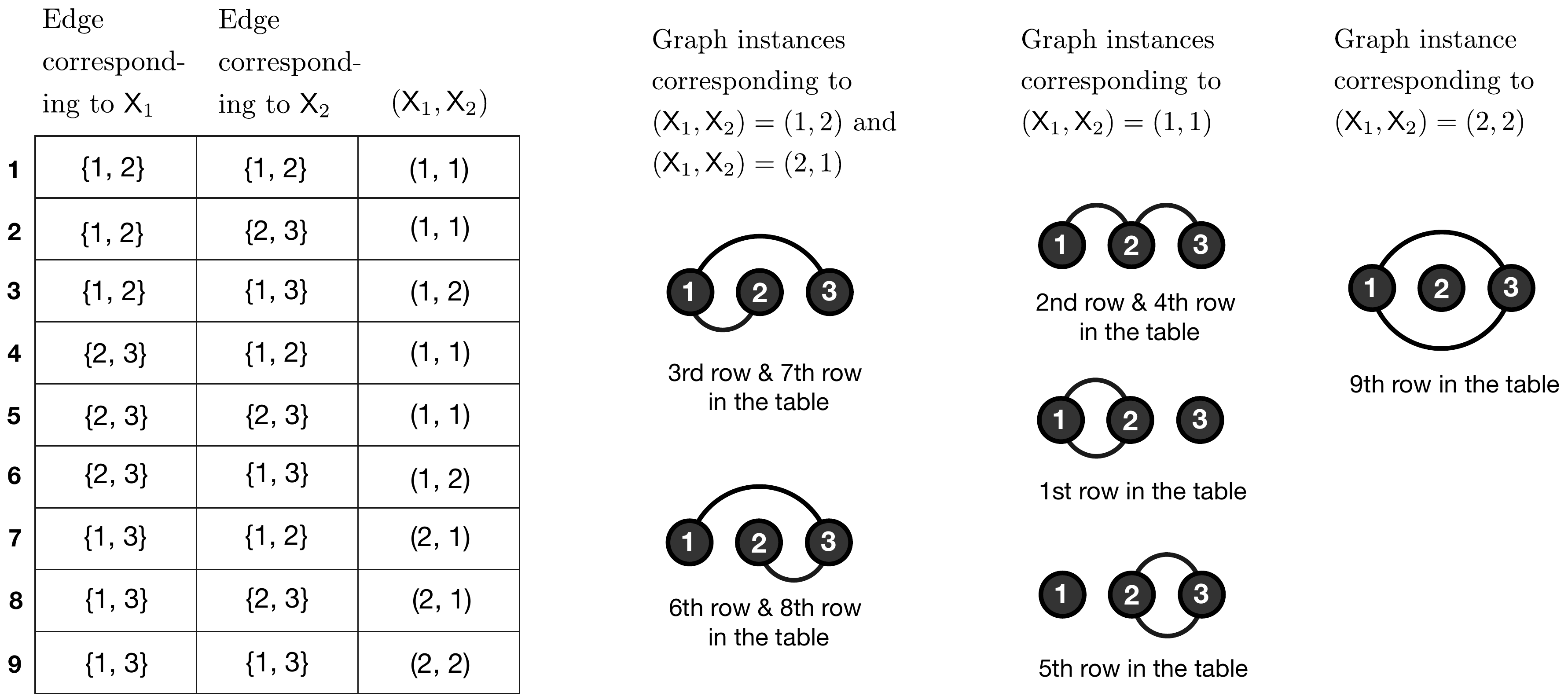}
  \caption{
Illustration of the possible patterns of $(\mathsf{X}_{1}, \mathsf{X}_{2})$ (left) for the graph ensemble with $N=3$ and $M=2$ and the corresponding graph instances (right). 
Each element in $(\mathsf{X}_{1}, \mathsf{X}_{2})$ is an outcome of the random variable defined by Eq.~(\ref{TriangularDistribution}). 
The vertex sequence is fixed in this example. 
	}
  \label{fig:XmStatisticExample}
\end{figure*}

We illustrate how the statistics with $\mathsf{H}_{1}$ random variable differs from those with $\stat{H}_{1}(\mat{A}; \ket{\pi})$ using a small graph. 
In Fig.~\ref{fig:XmStatisticExample}, we consider graphs with $N=3$ and $M=2$. 
The table on the left-hand side of Fig.~\ref{fig:XmStatisticExample} shows the possible outcomes of $(\mathsf{X}_{1}, \mathsf{X}_{2})$ and the corresponding edges for each element. 
Graph instances corresponding to each row are shown on the right. 
Herein, we assume that the vertex sequence is fixed; the sequence coincides with the labels on vertices. 
$\stat{H}_{1}(\mat{A}; \ket{\pi})$ is determined based on graph instances, whereas $\mathsf{H}_{1}$ is determined based on $(\mathsf{X}_{1}, \mathsf{X}_{2})$. 

The ordering of the edges matters in $(\mathsf{X}_{1}, \mathsf{X}_{2})$. 
For example, there are two graphs corresponding to $(\mathsf{X}_{1}, \mathsf{X}_{2}) = (1,2)$ or $(\mathsf{X}_{1}, \mathsf{X}_{2}) = (2,1)$. 
However, according to Eq.~(\ref{TriangularDistribution}), the probability of $(\mathsf{X}_{1}, \mathsf{X}_{2}) = (1,2)$ or $(\mathsf{X}_{1}, \mathsf{X}_{2}) = (2,1)$ is $2/3\times1/3 + 2/3\times1/3 = 4/9$, where the factor $4$ indicates the four realizations in the table on the left-hand size of Fig.~\ref{fig:XmStatisticExample} (third, sixth, seventh, and eighth rows). 
That is, each of the two graphs corresponds to two realizations representing different edge orderings. 
Similarly, the second and fourth rows in the table, which have $(\mathsf{X}_{1}, \mathsf{X}_{2}) = (1,1)$, correspond to the same graph. 
In contrast, for the other graphs instances, each graph corresponds to only one realization because the graph consists of a multiedge. 
This example illustrates that, whereas each simple graph has $M!$ realizations in $(\mathsf{X}_{1}, \dots, \mathsf{X}_{M})$, there are fewer than $M!$ realizations corresponding to a multigraph.

\section{Exact probability distribution of a test statistic under the Erd\H{o}s-R\'{e}nyi random graph model}\label{sec:ExactERrandom-Appendix}
In the main text, we employed approximate probability distributions of the $\stat{H}_{1}$ test statistics for the {\ER} model. 
The treatment in the main text was not exact because the number of multigraphs was not counted exactly. 
Here, we derive the exact probability distribution in which we accurately count the number of multigraphs. 
In this appendix, we consider an arbitrary affinity matrix $\mat{J}$. 
We then show how the exact distribution becomes approximately equivalent to that considered in the main text. 

We first count the total number of graphs with a fixed number of edges $M$.  
Using an integral representation of the Kronecker delta and the residue theorem, we obtain the following: 
\begin{align}
&\sum_{\{A_{ij}\}_{i<j}} \delta\left( M, \sum_{i<j} A_{ij} \right) \notag\\
&= \oint \frac{dz}{2\pi i} z^{-(1+M)} \prod_{i<j} \left( \sum_{A_{ij}=0}^{\infty} z^{A_{ij}} \right) \notag\\
&= \oint \frac{dz}{2\pi i} \frac{1}{z^{1+M}} \frac{1}{(1-z)^{\binom{N}{2}}} 
= \frac{1}{M!} \left. \frac{d^{M}}{dz^{M}} (1-z)^{-\binom{N}{2}} \right|_{z=0} \notag\\
&= \nHr{\binom{N}{2}}{M}. \label{ERGraphNumber}
\end{align}
Because all the allowed graph instances occur with equal probability, the probability with respect to the adjacency matrix is 
\begin{align}
\mathrm{Prob}\left[ \mat{A} \right] = \frac{ \delta \biggl( M, \sum_{i<j} A_{ij} \biggr) 
\displaystyle\prod_{i=1}^{N} \delta\left( A_{ii}, 0 \right) \prod_{i<j} \delta\left( A_{ij}, A_{ji} \right) }{\nHr{\binom{N}{2}}{M}}.
\end{align}
Thus, 
\begin{align}
&\mathrm{Prob}\left[ \stat{H}_{J} = E \right] \notag\\
&= \sum_{\mat{A}} \mathrm{Prob}\left[ \mat{A} \right] \delta\biggl( \beta_{\mat{J}} E, \sum_{i<j} J_{\pi_{i} \pi_{j}} A_{ij} \biggr) \notag\\
&= \frac{ \displaystyle\sum_{\{A_{ij}\}_{i<j}} \delta\biggl( M, \sum_{i<j} A_{ij} \biggr) 
\delta\biggl( \beta_{\mat{J}} E, \sum_{i<j} J_{\pi_{i} \pi_{j}} A_{ij} \biggr)}{\nHr{\binom{N}{2}}{M}} \label{ERExactDistribution1}
\end{align}
is the exact probability for $\stat{H}_{J}$. 
Here, the choice of the vertex sequence $\ket{\pi}$ does not really matter because we take the sum over all possible combinations of $\{A_{ij}\}_{i<j}$, and we can replace $J_{\pi_{i} \pi_{j}}$ with $J_{ij}$. 

Next, we investigate how the exact distribution (\ref{ERExactDistribution1}) is related to the approximate distribution. 
Analogous to the calculations in Eq.~(\ref{ERGraphNumber}), Eq.~(\ref{ERExactDistribution1}) can be modified as follows: 
\begin{align}
\mathrm{Prob}\left[ \stat{H}_{J} = E \right] 
&= \frac{1}{\nHr{\binom{N}{2}}{M}} 
\oint \frac{d\tilde{z}}{2\pi i} \tilde{z}^{-(1+\beta_{\mat{J}}E)} 
\oint \frac{dz}{2\pi i} z^{-(1+M)} \notag\\ 
&\hspace{20pt}\times \prod_{i<j} \left( \sum_{A_{ij}=0}^{\infty} \tilde{z}^{J_{ij}A_{ij}} z^{A_{ij}} \right) \notag\\
&= \frac{1}{\nHr{\binom{N}{2}}{M}} 
\oint \frac{d\tilde{z}}{2\pi i} \tilde{z}^{-(1+\beta_{\mat{J}}E)} 
\oint \frac{dz}{2\pi i} z^{-(1+M)} \notag\\
&\hspace{20pt}\times \exp\left( -\sum_{i<j} \log\left( 1 - \tilde{z}^{J_{ij}} z \right) \right). 
\label{ERExactDistribution2}
\end{align}
The complex integral can be computed with respect to $\tilde{z}$ by applying the residue theorem and conducting the $M$th-order derivative of an exponential function (the Fa\`{a} di Bruno's formula):
\begin{align}
\frac{d^{M}}{dz^{M}} e^{f(z)} 
&= e^{f(z)} \sum_{\{ n_{k} \}} \frac{M!}{\prod_{k=1}^{M} n_{k}! k!^{n_{k}}} \prod_{k=1}^{M} \left( \frac{d^{k} f(z) }{dz^{k}} \right)^{n_{k}}, \label{FaadiBruno}
\end{align}
where $\{ n_{k} \}$ indicates the set 
\begin{align}
& \{ n_{k} \} = \left\{ n_{k} \left| \sum_{k=1}^{M} k n_{k} = M \right.\right\}. 
\end{align}

Here, we do not employ the expansion in Eq.~(\ref{FaadiBruno}) although it is exact. 
Instead, we expand the logarithm up to the first order in $z$ in Eq.~(\ref{ERExactDistribution2}). 
This approximation allows us to clarify the difference between the exact and approximate distributions. 
Then, 
\begin{align}
&\mathrm{Prob}\left[ \stat{H}_{J} = E \right] \notag\\
&\approx \frac{1}{\nHr{\binom{N}{2}}{M}} 
\oint \frac{d\tilde{z}}{2\pi i} \tilde{z}^{-(1+\beta_{\mat{J}}E)} \notag\\
&\hspace{20pt}\times\left. \frac{1}{M!} \frac{\partial^{M}}{\partial z^{M}} 
\exp\left( z \sum_{i<j} \tilde{z}^{J_{ij}} \right) \right|_{z=0} \notag\\
&= \frac{1}{\nHr{\binom{N}{2}}{M}} \frac{1}{M!} 
\oint \frac{d\tilde{z}}{2\pi i} \tilde{z}^{-(1+\beta_{\mat{J}}E)} 
\left( \sum_{i<j} \tilde{z}^{J_{ij}} \right)^{M}. \label{ERExactDistribution3}
\end{align}
Here, we let the population of the affinity matrix elements be $P(\mathsf{J})$. 
Then, we have 
\begin{align}
\sum_{i<j} \tilde{z}^{J_{ij}} = \binom{N}{2} \sum_{\mathsf{J}=0}^{\infty} \tilde{z}^{\mathsf{J}} P(\mathsf{J}). 
\end{align}
Therefore, 
\begin{align}
&\mathrm{Prob}\left[ \stat{H}_{J} = E \right] \notag\\
&= \frac{\binom{N}{2}^{M}}{\nHr{\binom{N}{2}}{M} M!} 
\oint \frac{d\tilde{z}}{2\pi i} \tilde{z}^{-(1+\beta_{\mat{J}}E)} 
\sum_{\{\mathsf{J}_{m}\}} \tilde{z}^{\sum_{m} \mathsf{J}_{m}} \prod_{m=1}^{M} P(\mathsf{J}_{m}) \notag\\
&= \frac{\binom{N}{2}^{M}}{\nHr{\binom{N}{2}}{M} M!} 
\sum_{\{\mathsf{J}_{m}\}} \delta\left( \beta_{\mat{J}}E, \sum_{m} \mathsf{J}_{m}\right) \prod_{m=1}^{M} P(\mathsf{J}_{m}). 
\label{ERExactDistribution4}
\end{align}
This is equivalent to Eq.~(\ref{H1NullDistributionExact}), except for the normalization factor. 

The normalization is violated because of the approximation, that is, Eq.~(\ref{ERExactDistribution4}) is no longer a probability distribution. 
The amount of violation indicates how much we overcount the number of graphs; as we mentioned in the main text, we enumerate all possible sequences of edges including its order, which is $(\binom{N}{2})^{M}$, and correct it by $M!$. 
When the graph is sparse ($M = O(N)$), the amount of overcounting is of a constant order: 
\begin{align}
\frac{\binom{N}{2}^{M}}{M!} \left/ \nHr{\binom{N}{2}}{M} \right.
&= \exp\left( \sum_{k=0}^{M-1} \log \left( 1 + \frac{k}{\binom{N}{2}} \right) \right) \notag\\
&\approx \exp\left( \frac{1}{\binom{N}{2}} \sum_{k=0}^{M-1} k \right) \notag\\
&= e^{\frac{M(M-1)}{N(N-1)}}
= O(1). 
\end{align}
The fact that the violation of the overall normalization is $O(1)$ implies that the deviation of the probability $\mathrm{Prob}\left[ \stat{H}_{J} \right]$ at each point becomes negligibly small as the data size increases. 
In other words, when graphs are sufficiently dense, the effect of multiedge indistinguishability can significantly contribute to the null distribution.

\section{Probability distribution of a test statistic under the Erd\H{o}s-R\'{e}nyi random graph model with the canonical constraint}\label{sec:CanonicalERrandom-Appendix}
In this appendix, as a variant of the {\ER} model considered in the main text, we consider the model in which the adjacency matrix elements are generated independently as follows: 
\begin{align}
\mathrm{Prob} \left[ \mat{A} \right] 
= \prod_{i=1}^{N} \delta\left( A_{ii}, 0 \right) \prod_{i < j} \frac{ \lambda^{A_{ij}} }{A_{ij}!} e^{-\lambda} \delta\left( A_{ij}, A_{ji} \right), \label{CanonicalER1-Appendix}
\end{align}
where $\lambda$ is determined such that the total number of edges coincides with the observed value $M$ on average, that is, 
\begin{align}
\mathbb{E}_{\mat{A}}\left[ \sum_{i<j} A_{ij}\right] = \lambda \binom{N}{2} = M.
\end{align}
This is often referred to as the canonical constraint. 

Using the tricks used in {\Methods}~\ref{sec:ExactERrandom-Appendix}, we obtain the probability distribution for $\stat{H}_{J}$ in terms of the population of the affinity matrix elements $P(\mathsf{J})$ as follows: 
\begin{align}
&\mathrm{Prob}\left[ \stat{H}_{J} = E \right] \notag\\
&= \sum_{\{A_{ij}\}_{i<j}} 
\delta\biggl( \beta_{\mat{J}} E, \sum_{i<j} J_{\pi_{i} \pi_{j}} A_{ij} \biggr) 
\prod_{i < j} \frac{ \lambda^{A_{ij}} }{A_{ij}!} e^{-\lambda} \notag\\
&= e^{-\lambda \binom{N}{2}} \oint \frac{dz}{2\pi i} z^{-(1+\beta_{\mat{J}}E)} 
\prod_{i<j} \left( \sum_{A_{ij}=0}^{\infty} \frac{ \lambda^{A_{ij}} }{A_{ij}!} z^{J_{\pi_{i} \pi_{j}} A_{ij}} \right) \notag\\
&= e^{-\lambda \binom{N}{2}} \oint \frac{dz}{2\pi i} z^{-(1+\beta_{\mat{J}}E)} \exp\left( \lambda \sum_{i<j}z^{J_{ij}} \right) \notag\\
&= e^{-\lambda \binom{N}{2}} \oint \frac{dz}{2\pi i} z^{-(1+\beta_{\mat{J}}E)} \exp\left( \lambda \binom{N}{2} \sum_{\mathsf{J}} z^{\mathsf{J}} P(\mathsf{J}) \right) \notag\\
&= e^{-\lambda \binom{N}{2}} \sum_{k=0}^{\infty} \frac{\left(\lambda \binom{N}{2}\right)^{k}}{k!} 
\sum_{ \{\mathsf{J}_{m}\} } \delta\biggl( \beta_{\mat{J}}E, \sum_{m=1}^{k} \mathsf{J}_{m} \biggr) \prod_{m=0}^{k} P\left(\mathsf{J}_{m}\right) \notag\\
&= \sum_{k=0}^{\infty} \mathrm{Poi}\left( k; M \right) \sum_{ \{\mathsf{J}_{m}\} } \delta\biggl( \beta_{\mat{J}}E, \sum_{m=1}^{k} \mathsf{J}_{m} \biggr) \prod_{m=0}^{k} P\left(\mathsf{J}_{m}\right). \label{CanonicalER2-Appendix}
\end{align}
Here, $\mathrm{Poi}\left( k; M \right)$ is the Poisson distribution with respect to $k$ with mean $M$. 
Note that the latter half of Eq.~(\ref{CanonicalER2-Appendix}) is analogous to the null distributions considered in the main text. 

Because the Poisson distribution is highly peaked around its mean when the mean value is sufficiently large, we have 
\begin{align}
\mathrm{Prob}\left[ \stat{H}_{J} = E \right] 
\simeq 
\sum_{ \{\mathsf{J}_{m}\} } \delta\biggl( \beta_{\mat{J}}E, \sum_{m=1}^{M} \mathsf{J}_{m} \biggr) \prod_{m=0}^{M} P\left(\mathsf{J}_{m}\right). \label{CanonicalER3-Appendix}
\end{align}
Therefore, when $M$ is large, the null hypothesis in this appendix yields the same null distributions as Eq.~(\ref{H1NullDistributionExact}). 
Note that this is also the case in which the normal approximation is accurate because of the central limit theorem.

\section{Fallacy of statistical test for optimized sequences} \label{sec:NullHypothesisFallacy}
It is difficult to formulate a hypothesis testing in which we consider adjacency matrices with optimized vertex sequences as a null hypothesis. 
Here, we explain the reasons for this with some specific examples. 

Even when graphs are generated from a random graph model, the adjacency matrices can exhibit strong locality structures when an envelope reduction algorithm is executed on each generated graph (Fig.~\ref{fig:Optimized_ERgraphs}). 
Thus, the resulting $\stat{H}_{J}$ test statistics corresponding to the optimized adjacency matrices may sensitively depend on the tuning parameters of the optimization algorithm. 
Different null distributions may be obtained depending on the initial condition of an iterative algorithm because the algorithm may converge to different local optima. 
Moreover, the test statistic $\stat{H}_{J}$ may no longer be expressed as a sum of identically distributed random variables, indicating that there is no guarantee that the distribution is approximately normal. 
This is easy to imagine, for example, if we consider $\stat{H}_{J}$ written as a function of matrix eigenvalues and recall that a random matrix often has a ``semi-circle law'' as a limiting eigenvalue distribution.
Therefore, we cannot naively use the built-in standard error of a statistical analysis package in which a normal distribution is assumed. 

An empirical distribution of the test statistic can be obtained by generating synthetic graphs and performing optimization on each of them. 
One might be tempted to use this empirical distribution (or its bootstrap distribution) to draw the standard error and compute the $p$ value. 
However, we still have the problem of algorithmic uncertainty. Even if the observed dataset is identified as having a significant sequentially local structure based on the empirical null distribution, we cannot conclude whether it is because of the dataset itself, or because of the choice of the algorithm and its tuning parameters. 
However, one can assess the significance of a graph without being affected by algorithmic uncertainties when one uses, for instance, a deterministic algorithm without tuning parameters or an algorithm that always yields a unique solution. Nevertheless, even in these cases, one needs to keep in mind that the statistical assessment is conditioned on the algorithm employed. 

In summary, the crucial issue in using optimized adjacency matrices as a null hypothesis is that the null distribution of the test statistic is generally affected by the optimization algorithm in a nontrivial way.
As a result, one does not exactly know what is really assumed as the null hypothesis and, accordingly, how to interpret the obtained $p$ value. 
To conduct a statistical test of sequential locality, therefore, we have to have an interpretable null hypothesis, such as that based on the ORGM.

\section{Distribution of the test statistic in the ordered random graph model \label{sec:SimplegraphORGM}}
We derive the $\stat{H}_{1}$-statistic distribution when the graphs are generated from the ORGM, which we denote as $\mathrm{Prob}\left[ \stat{H}_{1} = E; N, M, r, \epsilon \right]$. 
We assume that the vertex sequence $\ket{\pi}^{\ast}$ is aligned in the intrinsic order of the ORGM. 
Herein, we focus on the ORGM that is constrained to simple graphs.
The ORGM without this constraint is analyzed in {\Methods}~\ref{sec:MultigraphORGM}. 

The first moment of an adjacency matrix element with respect to Eq.~(\ref{ORGM-Prob}) is 
\begin{align}
\mathbb{E}_{\mat{A}}\left[ A_{ij} \right] 
&= \begin{cases}
\displaystyle \frac{\binom{\left|\Omega_{\mathrm{in}}\right|-1}{M_{\mathrm{in}}-1}}{\binom{\left|\Omega_{\mathrm{in}}\right|}{M_{\mathrm{in}}}} = \frac{M_{\mathrm{in}}}{\left|\Omega_{\mathrm{in}}\right|} & \text{for } (i,j) \in \Omega_{\mathrm{in}} \\[15pt]
\displaystyle \frac{\binom{\left|\Omega_{\mathrm{out}}\right|-1}{M_{\mathrm{out}}-1}}{\binom{\left|\Omega_{\mathrm{out}}\right|}{M_{\mathrm{out}}}} = \frac{M_{\mathrm{out}}}{\left|\Omega_{\mathrm{out}}\right|} & \text{for } (i,j) \in \Omega_{\mathrm{out}} 
\end{cases}
\end{align}
where the numerator represents the number of allowed adjacency matrices given that $A_{ij}=1$. 
The second moment is obtained similarly as 
\begin{widetext}
\begin{align}
&\mathbb{E}_{\mat{A}}\left[ A_{ij} A_{k\ell} \right] 
= \begin{cases}
\displaystyle \frac{\binom{\left|\Omega_{\mathrm{in}}\right|-1}{M_{\mathrm{in}}-1}}{\binom{\left|\Omega_{\mathrm{in}}\right|}{M_{\mathrm{in}}}} = \frac{M_{\mathrm{in}}}{\left|\Omega_{\mathrm{in}}\right|} & \text{for } (i,j)=(k,\ell) \in \Omega_{\mathrm{in}} \\[15pt]
\displaystyle \frac{\binom{\left|\Omega_{\mathrm{out}}\right|-1}{M_{\mathrm{out}}-1}}{\binom{\left|\Omega_{\mathrm{out}}\right|}{M_{\mathrm{out}}}} = \frac{M_{\mathrm{out}}}{\left|\Omega_{\mathrm{out}}\right|} & \text{for } (i,j)=(k,\ell) \in \Omega_{\mathrm{out}} \\[15pt]
\displaystyle \frac{\binom{\left|\Omega_{\mathrm{in}}\right|-2}{M_{\mathrm{in}}-2}}{\binom{\left|\Omega_{\mathrm{in}}\right|}{M_{\mathrm{in}}}} = \frac{M_{\mathrm{in}} \left( M_{\mathrm{in}}-1 \right)}{\left|\Omega_{\mathrm{in}}\right| \left( \left|\Omega_{\mathrm{in}}\right|-1 \right)} & \text{for } (i,j)\neq(k,\ell) \in \Omega_{\mathrm{in}} \\[15pt]
\displaystyle \frac{\binom{\left|\Omega_{\mathrm{out}}\right|-2}{M_{\mathrm{out}}-2}}{\binom{\left|\Omega_{\mathrm{out}}\right|}{M_{\mathrm{out}}}} = \frac{M_{\mathrm{out}} \left( M_{\mathrm{out}}-1 \right)}{\left|\Omega_{\mathrm{out}}\right| \left( \left|\Omega_{\mathrm{out}}\right|-1 \right)} & \text{for } (i,j)\neq(k,\ell) \in \Omega_{\mathrm{out}} \\[15pt]
\displaystyle \frac{\binom{\left|\Omega_{\mathrm{in}}\right|-1}{M_{\mathrm{in}}-1}}{\binom{\left|\Omega_{\mathrm{in}}\right|}{M_{\mathrm{in}}}} 
\frac{\binom{\left|\Omega_{\mathrm{out}}\right|-1}{M_{\mathrm{out}}-1}}{\binom{\left|\Omega_{\mathrm{out}}\right|}{M_{\mathrm{out}}}} 
= \frac{M_{\mathrm{in}}}{\left|\Omega_{\mathrm{in}}\right|} \frac{M_{\mathrm{out}}}{\left|\Omega_{\mathrm{out}}\right|} 
&\text{for }\left\{ 
\begin{aligned}
    (i,j) \in \Omega_{\mathrm{in}},\, (k,\ell) \in \Omega_{\mathrm{out}} \\
    (i,j) \in \Omega_{\mathrm{out}},\, (k,\ell) \in \Omega_{\mathrm{in}}
\end{aligned}  \right.
\end{cases}.
\end{align}
Using these moments, the first and second moments of the $\stat{H}_{1}$ statistic are 
\begin{align}
\mathbb{E}_{\mat{A}}\left[ \stat{H}_{1}(\mat{A}, \ket{\pi}^{\ast}) \right] 
&= \frac{1}{\beta_{1}}\sum_{i<j} \mathbb{E}_{\mat{A}}\left[ A_{ij} \right] | i-j | \notag\\
&= \frac{1}{\beta_{1}} \frac{M_{\mathrm{in}}}{\left|\Omega_{\mathrm{in}}\right|} 
\frac{r(r+1) (3N - 2r - 1)}{6}
+ \frac{1}{\beta_{1}} \frac{M_{\mathrm{out}}}{\left|\Omega_{\mathrm{out}}\right|} 
\frac{N^{3} - N(3r^{2}+3r+1) + r(2r^{2}+3r+1)}{6} \label{ORGM-1stMoment}
\end{align}
and 
\begin{align}
&\mathbb{E}_{\mat{A}}\left[ \stat{H}^{2}_{1}(\mat{A}, \ket{\pi}^{\ast}) \right] 
= \frac{1}{\beta^{2}_{1}} \sum_{i<j} \sum_{k<\ell} \mathbb{E}_{\mat{A}}\left[ A_{ij} A_{k\ell} \right] 
\left|\pi_{i} - \pi_{j}\right| \left|\pi_{k} - \pi_{\ell}\right| \notag\\
&= \frac{1}{\beta^{2}_{1}} \frac{M_{\mathrm{in}}}{\left|\Omega_{\mathrm{in}}\right|} 
\frac{r^{2}(r+1)^{2}}{6} \left( \frac{N (2r+1)}{r(r+1)} - \frac{3}{2} \right) 
+ \frac{1}{\beta^{2}_{1}} \frac{M_{\mathrm{out}}}{\left|\Omega_{\mathrm{out}}\right|} \frac{(N-r)(N-r-1)}{12} 
\left( \left( N+r+\frac{1}{2} \right)^{2} + 2r(r+1) -\frac{1}{4} \right) \notag\\
&\hspace{10pt}+ \frac{1}{\beta^{2}_{1}} \frac{M_{\mathrm{in}} (M_{\mathrm{in}}-1)}{\left|\Omega_{\mathrm{in}}\right| (\left|\Omega_{\mathrm{in}}\right|-1)} 
\frac{r^{2}(r+1)^{2}}{6} \left( 
\frac{\left( 3N-2r-1 \right)^{2}}{6} - \frac{N (2r+1)}{r(r+1)} + \frac{3}{2}
\right) \notag\\
&\hspace{10pt}+\frac{1}{\beta^{2}_{1}} \frac{M_{\mathrm{out}} (M_{\mathrm{out}}-1)}{\left|\Omega_{\mathrm{out}}\right| (\left|\Omega_{\mathrm{out}}\right|-1)} 
\frac{(N-r) (N+2r) (N-r-1) (N-r+1) (N-r-2) (N+2r+2)}{36} \notag\\
&\hspace{10pt}+ \frac{2}{\beta^{2}_{1}} \frac{M_{\mathrm{in}} M_{\mathrm{out}}}{\left|\Omega_{\mathrm{in}}\right| \left|\Omega_{\mathrm{out}}\right|} 
\frac{r(r+1) (N-r) (N-r-1) (N+2r+1) \left(3N-2r-1 \right)}{36}. \label{ORGM-2ndMoment}
\end{align}
\end{widetext}
Then, the variance of the $\stat{H}_{1}$ statistic is given by 
$\mathrm{Var}_{\mat{A}}\left[ \stat{H}_{1}(\mat{A}, \ket{\pi}) \right]
= \mathbb{E}_{\mat{A}}\left[ \stat{H}^{2}_{1}(\mat{A}, \ket{\pi}) \right] 
- \mathbb{E}_{\mat{A}}\left[ \stat{H}_{1}(\mat{A}, \ket{\pi}) \right]^{2}$. 
In principle, we can also compute higher-order moments analogously. 

Note that we cannot apply the central limit theorem to Eq.~(\ref{ORGM-Prob}) and obtain the asymptotic distribution of $\stat{H}_{1}$. 
This is because the edge generation processes are not independent of each other, as the numbers of edges $M_{\mathrm{in}}$ and $M_{\mathrm{out}}$ are strictly constrained, and multiedges are not allowed. 
However, note also that the ORGM is a compound model that consists of an {\ER} graph in each of $\Omega_{\mathrm{in}}$ and $\Omega_{\mathrm{out}}$, and recall that the test-statistic distribution for the {\ER} model is approximately normal in many cases. 
Therefore, unless the graph size is very small and/or dense, we can expect that $\mathrm{Prob}\left[ \stat{H}_{1} = E; N, M, r, \epsilon \right]$ is approximately normal, i.e., 
\begin{align}
&\mathrm{Prob}\left[ \stat{H}_{1} = E; N, M, r, \epsilon \right] \notag\\
&\hspace{10pt}\approx \mathcal{N}\left(\mathbb{E}_{\mat{A}}\left[ \stat{H}_{1}(\mat{A}, \ket{\pi}) \right], \sqrt{\mathrm{Var}_{\mat{A}}\left[ \stat{H}_{1}(\mat{A}, \ket{\pi}) \right]}  \right). \label{ORGMNormalDistribution}
\end{align}

\section{Maximum likelihood estimate of the ordered random graph model}\label{sec:MLE_r}
We derive the MLEs of the ORGM parameters. 
We first consider the bandwidth $r$, which determines $\Omega_{\mathrm{in}}$ and $\Omega_{\mathrm{out}}$. 
Given an adjacency matrix with a specified vertex sequence, the bandwidth automatically determines $M_{\mathrm{in}}$ and $M_{\mathrm{out}}$. 
For the simple-graph variant of the ORGM, the maximizer $r^{\ast}$ of the log-likelihood function corresponding to Eq.~(\ref{ORGM-Prob}) is 
\begin{align}
r^{\ast} = \argmin_{r} \left\{ \log \binom{\left|\Omega_{\mathrm{in}}\right|}{M_{\mathrm{in}}} + \log\binom{\left|\Omega_{\mathrm{out}}\right|}{M_{\mathrm{out}}} \right\}. \label{MLE-r}
\end{align}
In the actual implementation, we evaluate the microcanonical entropies in Eq.~(\ref{MLE-r}) using Stirling's approximation, as we sweep $r$. 
Given the MLE of the bandwidth $r^{\ast}$, the density ratio $\epsilon^{\ast}$ with nonzero likelihood is uniquely determined as 
\begin{align}
\epsilon^{\ast} = \frac{M^{\ast}_{\mathrm{out}}/\left|\Omega^{\ast}_{\mathrm{out}}\right|}{M^{\ast}_{\mathrm{in}}/\left|\Omega^{\ast}_{\mathrm{in}}\right|}, 
\end{align}
where $\left|\Omega^{\ast}_{\mathrm{in}}\right|$, $M^{\ast}_{\mathrm{in}}$, $\left|\Omega^{\ast}_{\mathrm{out}}\right|$, and $M^{\ast}_{\mathrm{out}}$ are the MLEs corresponding to $r^{\ast}$.

\section{Ordered random graph model allowing multiedges \label{sec:MultigraphORGM}}
We show that the results in {\Methods}~\ref{sec:SimplegraphORGM} are altered when graphs are allowed to have multiedges in the ORGM. 
The first moment of adjacency matrix elements with respect to Eq.~(\ref{ORGM-Prob}) is 
\begin{align}
\mathbb{E}_{\mat{A}}\left[ A_{ij} \right] 
&= \begin{cases}
\displaystyle \frac{ \nHr{\left|\Omega_{\mathrm{in}}\right|+1}{M_{\mathrm{in}}-1} }{ \nHr{\left|\Omega_{\mathrm{in}}\right|}{M_{\mathrm{in}}} } 
= \frac{M_{\mathrm{in}}}{\left|\Omega_{\mathrm{in}}\right|} & \text{for } (i,j) \in \Omega_{\mathrm{in}} \\[15pt]
\displaystyle \frac{ \nHr{\left|\Omega_{\mathrm{out}}\right|+1}{M_{\mathrm{out}}-1} }{ \nHr{\left|\Omega_{\mathrm{out}}\right|}{M_{\mathrm{out}}} } 
= \frac{M_{\mathrm{out}}}{\left|\Omega_{\mathrm{out}}\right|} & \text{for } (i,j) \in \Omega_{\mathrm{out}} 
\end{cases}. \label{multigraphORGM-1stMoment}
\end{align}
Hence, the first moment is identical to the case with the simple-graph constraint. 
Here, we obtained the numerator as follows. 
For any $(k,\ell) \in \Omega_{\mathrm{in}}$, 
\begin{align}
&\sum_{\{ A_{ij} | (i,j) \in \Omega_{\mathrm{in}} \}} 
\delta\left( M_{\mathrm{in}}, \sum_{(i,j) \in \Omega_{\mathrm{in}}} A_{ij} \right) A_{k\ell} \notag\\
&= \sum_{\{ A_{ij} | (i,j) \in \Omega_{\mathrm{in}} \}} 
\oint \frac{dz}{2\pi i} z^{\sum_{(i,j) \in \Omega_{\mathrm{in}}} A_{ij} - M_{\mathrm{in}} - 1} A_{k\ell} \notag\\
&= \oint \frac{dz}{2\pi i} \frac{1}{z^{1+M_{\mathrm{in}}}}
\left( \sum_{ A_{k\ell}=0 }^{\infty} A_{k\ell} \, z^{A_{k\ell}} \right) 
\prod_{ \substack{(i,j) \in \Omega_{\mathrm{in}} \\ (i,j) \ne (k,\ell)} } \left( \sum_{ A_{ij}=0 }^{\infty} z^{A_{ij}} \right) \notag\\
&= \oint \frac{dz}{2\pi i} 
\frac{1}{z^{M_{\mathrm{in}}} (1-z)^{1+|\Omega_{\mathrm{in}}|}} \notag\\
&= \frac{1}{(M_{\mathrm{in}}-1)!} \left.\frac{d^{M_{\mathrm{in}}-1}}{dz^{M_{\mathrm{in}}-1}} \frac{1}{(1-z)^{1+|\Omega_{\mathrm{in}}|}}\right|_{z=0} \notag\\
&= \nHr{\left|\Omega_{\mathrm{in}}\right|+1}{M_{\mathrm{in}}-1}, \label{IntegralTrick1}
\end{align}
where $\oint dz$ is a complex integral along a closed path around $z=0$, which does not contain $z=1$ inside. 
Here, we used an integral representation of the Kronecker delta (z-transform) and the residue theorem. 
To interpret this quantity, it should be noted that $\nHr{\left|\Omega_{\mathrm{in}}\right|+1}{M_{\mathrm{in}}-1} = \nHr{\left|\Omega_{\mathrm{in}}\right|}{M_{\mathrm{in}}} \times (\left|\Omega_{\mathrm{in}}\right|/M_{\mathrm{in}})$; among the $\nHr{\left|\Omega_{\mathrm{in}}\right|}{M_{\mathrm{in}}}$ allowed matrices, the average value of any matrix element is given by $\left|\Omega_{\mathrm{in}}\right|/M_{\mathrm{in}}$. 

The quantities in Eqs.~(\ref{ORGM-1stMoment}), (\ref{ORGM-2ndMoment}), and (\ref{multigraphORGM-1stMoment}) can be obtained through the classical approach in combinatorics, that is, by directly counting the number of possible outcomes. 
However, for more complicated and less intuitive quantities, such as the second moments of adjacency matrix elements in multigraphs, the classical approach becomes increasingly difficult. 
In such cases, the trick which offers a systematic prescription, as shown in Eq.~(\ref{IntegralTrick1}), becomes more valuable; for example, the numerator of $\mathbb{E}_{\mat{A}}\left[ A^{2}_{k\ell} \right]$, where $(k,\ell) \in \Omega_{\mathrm{in}}$ is calculated as 
\begin{align}
&\sum_{\{ A_{ij} | (i,j) \in \Omega_{\mathrm{in}} \}} 
\delta\left( M_{\mathrm{in}}, \sum_{(i,j) \in \Omega_{\mathrm{in}}} A_{ij} \right) A^{2}_{k\ell} \notag\\
&= \sum_{\{ A_{ij} | (i,j) \in \Omega_{\mathrm{in}} \}} 
\oint \frac{dz}{2\pi i} z^{\sum_{(i,j) \in \Omega_{\mathrm{in}}} A_{ij} - M_{\mathrm{in}} - 1} A^{2}_{k\ell} \notag\\
&= \oint \frac{dz}{2\pi i} \frac{1}{z^{1+M_{\mathrm{in}}}}
\left( \sum_{ A_{k\ell}=0 }^{\infty} A^{2}_{k\ell} \, z^{A_{k\ell}} \right) 
\prod_{ \substack{(i,j) \in \Omega_{\mathrm{in}} \\ (i,j) \ne (k,\ell)} } \left( \sum_{ A_{ij}=0 }^{\infty} z^{A_{ij}} \right) \notag\\
&= \oint \frac{dz}{2\pi i} 
\frac{1+z}{z^{M_{\mathrm{in}}} (1-z)^{2+|\Omega_{\mathrm{in}}|}} \notag\\
&= \frac{1}{(M_{\mathrm{in}}-1)!} \left.\frac{d^{M_{\mathrm{in}}-1}}{dz^{M_{\mathrm{in}}-1}} \frac{1+z}{(1-z)^{2+|\Omega_{\mathrm{in}}|}}\right|_{z=0} \notag\\
&= \nHr{\left|\Omega_{\mathrm{in}}\right|+2}{M_{\mathrm{in}}-1} 
+ \nHr{\left|\Omega_{\mathrm{in}}\right|+2}{M_{\mathrm{in}}-2}. \label{IntegralTrick2}
\end{align}
Here, we assumed that $M_{\mathrm{in}} > 1$. 
The second moments in other cases can also be calculated analogously. 

In summary, the second moments are 
\begin{widetext}
\begin{align}
&\mathbb{E}_{\mat{A}}\left[ A_{ij} A_{k\ell} \right] 
= \begin{cases}
\displaystyle \frac{ \nHr{\left|\Omega_{\mathrm{in}}\right|+2}{M_{\mathrm{in}}-1} + \nHr{\left|\Omega_{\mathrm{in}}\right|+2}{M_{\mathrm{in}}-2} }{ \nHr{\left|\Omega_{\mathrm{in}}\right|}{M_{\mathrm{in}}} } 
= \frac{M_{\mathrm{in}}}{\left|\Omega_{\mathrm{in}}\right|} 
\frac{\left|\Omega_{\mathrm{in}}\right|+2M_{\mathrm{in}} -1}{\left|\Omega_{\mathrm{in}}\right|+1} 
& \text{for } (i,j)=(k,\ell) \in \Omega_{\mathrm{in}} \\[15pt]
\displaystyle \frac{ \nHr{\left|\Omega_{\mathrm{out}}\right|+2}{M_{\mathrm{out}}-1} + \nHr{\left|\Omega_{\mathrm{out}}\right|+2}{M_{\mathrm{out}}-2} }
{ \nHr{\left|\Omega_{\mathrm{out}}\right|}{M_{\mathrm{out}}} } 
= \frac{M_{\mathrm{out}}}{\left|\Omega_{\mathrm{out}}\right|} 
\frac{\left|\Omega_{\mathrm{out}}\right|+2M_{\mathrm{out}} -1}{\left|\Omega_{\mathrm{out}}\right|+1}
& \text{for } (i,j)=(k,\ell) \in \Omega_{\mathrm{out}} \\[15pt]
\displaystyle \frac{ \nHr{\left|\Omega_{\mathrm{in}}\right|+2}{M_{\mathrm{in}}-2} }{ \nHr{\left|\Omega_{\mathrm{in}}\right|}{M_{\mathrm{in}}} } 
= \frac{M_{\mathrm{in}} \left( M_{\mathrm{in}}-1 \right)}{\left|\Omega_{\mathrm{in}}\right| \left( \left|\Omega_{\mathrm{in}}\right|+1 \right)} & \text{for } (i,j)\neq(k,\ell) \in \Omega_{\mathrm{in}} \\[15pt]
\displaystyle \frac{ \nHr{\left|\Omega_{\mathrm{out}}\right|+2}{M_{\mathrm{out}}-2} }{ \nHr{\left|\Omega_{\mathrm{out}}\right|}{M_{\mathrm{out}}} } 
= \frac{M_{\mathrm{out}} \left( M_{\mathrm{out}}-1 \right)}{\left|\Omega_{\mathrm{out}}\right| \left( \left|\Omega_{\mathrm{out}}\right|+1 \right)} & \text{for } (i,j)\neq(k,\ell) \in \Omega_{\mathrm{out}} \\[15pt]
\displaystyle 
\frac{ \nHr{\left|\Omega_{\mathrm{in}}\right|+1}{M_{\mathrm{in}}-1} }{ \nHr{\left|\Omega_{\mathrm{in}}\right|}{M_{\mathrm{in}}} }
\frac{ \nHr{\left|\Omega_{\mathrm{out}}\right|+1}{M_{\mathrm{out}}-1} }{ \nHr{\left|\Omega_{\mathrm{out}}\right|}{M_{\mathrm{out}}} }
= \frac{M_{\mathrm{in}}}{\left|\Omega_{\mathrm{in}}\right|} \frac{M_{\mathrm{out}}}{\left|\Omega_{\mathrm{out}}\right|} 
&\text{for }\left\{  
\begin{aligned}
    (i,j) \in \Omega_{\mathrm{in}},\, (k,\ell) \in \Omega_{\mathrm{out}} \\
    (i,j) \in \Omega_{\mathrm{out}},\, (k,\ell) \in \Omega_{\mathrm{in}}
\end{aligned}  \right.
\end{cases}. \label{multigraphORGM-2ndMoment}
\end{align}
\end{widetext}
We can obtain $\mathbb{E}_{\mat{A}}\left[ \stat{H}_{1}(\ket{\pi}^{\ast}; \mat{A}) \right]$ and $\mathbb{E}_{\mat{A}}\left[ \stat{H}^{2}_{1}(\ket{\pi}^{\ast}; \mat{A}) \right]$ by replacing the factors corresponding to $\mathbb{E}_{\mat{A}}\left[ A_{ij} \right]$ and $\mathbb{E}_{\mat{A}}\left[ A_{ij} A_{k\ell} \right]$ in Eqs.~(\ref{ORGM-1stMoment}) and (\ref{ORGM-2ndMoment}) with the values in Eqs.~(\ref{multigraphORGM-1stMoment}) and (\ref{multigraphORGM-2ndMoment}). 

These results indicate that the effect of the simple-graph constraint in the ORGM on our statistical test is not prominent when $\left|\Omega_{\mathrm{in}}\right|$ and $\left|\Omega_{\mathrm{out}}\right|$ are sufficiently large compared with $M_{\mathrm{in}}$ and $M_{\mathrm{out}}$, respectively. 
It should also be noted that the ORGM becomes equivalent to the {\ER} model when $\left|\Omega_{\mathrm{in}}\right|$ or $\left|\Omega_{\mathrm{out}}\right|$ coincides with all of the upper-right elements in an adjacency matrix, that is, the cases where $r = N-1$ or $r = 0$. 
Therefore, the results here also describe the distinction between the {\ER} models with and without the simple graph constraint in the $\stat{H}_{1}$ test statistics. 

Analogous to Eq.~(\ref{MLE-r}), the MLE $r^{\ast}$ of the bandwidth is obtained as 
\begin{align}
r^{\ast} = \argmin_{r} \left\{ \log\nHr{\left|\Omega_{\mathrm{in}}\right|}{M_{\mathrm{in}}} + \log\nHr{\left|\Omega_{\mathrm{out}}\right|}{M_{\mathrm{out}}} \right\}. \label{multigraphMLE-r}
\end{align}
In general, if $n \gg m$, then $\log\binom{n}{m} \approx m\log n - m^{2}/n$ and $\log\nHr{n}{m} \approx m\log n + m^{2}/n - 2m/n$; these are both dominated by $m\log n$. 
Hence, the MLE $r^{\ast}$ given by Eq.~(\ref{multigraphMLE-r}) is expected to be close to or coincide with that given by Eq.~(\ref{MLE-r}) when graphs are sparse.

\section{Test statistics under the random sequences \label{sec:RandomSequence-Appendix}}
In this appendix, we present the detailed derivation of Eqs.~(\ref{MeanRandomSequenceH1}) and (\ref{VarianceRandomSequenceH1}). 
In the following, we assume that $N > 3$.

\subsection{Mean of the $z_{1}$ statistic}
First, we calculate the ensemble average of $\stat{H}_{1}$ and $\stat{H}_{G}$ statistics with respect to the sequences. 
\begin{align}
\mathbb{E}_{\ket{\pi}}\left[ \stat{H}_{1}(\mat{A}, \ket{\pi}) \right] 
&= \frac{1}{\beta_{1}}\sum_{i<j} A_{ij} \mathbb{E}_{\ket{\pi}}\left[ \left|\pi_{i} - \pi_{j}\right| \right] \notag\\
&= \frac{1}{\beta_{1}} \frac{N+1}{3} \sum_{i<j} A_{ij}
= 1. 
\end{align}
Note that the average $\mathbb{E}_{\ket{\pi}}\left[ \left|\pi_{i} - \pi_{j}\right| \right]$ is equal to the average with respect to $\mathsf{X}$. 
The mean value of the $z_{1}$ statistic then reads 
\begin{align}
\mathbb{E}_{\ket{\pi}}\left[ z_{1}(\mat{A}, \ket{\pi}) \right] 
&= \sqrt{\frac{2M (N+1)}{N-2}} \left( \mathbb{E}_{\ket{\pi}}\left[ \stat{H}_{1}(\mat{A}, \ket{\pi}) \right] - 1 \right) 
= 0. 
\end{align}

\begin{widetext}
\subsection{Variance of the $z_{1}$ statistic}
We start with the second moment of the $\stat{H}_{1}$ statistic, 
\begin{align}
& \mathbb{E}_{\ket{\pi}}\left[ \stat{H}^{2}_{1}(\mat{A}, \ket{\pi}) \right] 
= \frac{1}{\beta^{2}_{1}}\sum_{i<j} A_{ij} \sum_{i^{\prime} < j^{\prime}} A_{i^{\prime} j^{\prime}} 
\mathbb{E}_{\ket{\pi}}\left[ \left|\pi_{i} - \pi_{j}\right| \left|\pi_{i^{\prime}} - \pi_{j^{\prime}}\right| \right]. \label{SecondMomentH1}
\end{align}
This is decomposed as follows: 
\begin{align}
\mathbb{E}_{\ket{\pi}}\left[ \stat{H}^{2}_{1}(\mat{A}, \ket{\pi}) \right] 
&= \frac{1}{\beta^{2}_{1}}
\sum_{i<j} A_{ij} \sum_{i^{\prime} < j^{\prime}} A_{i^{\prime} j^{\prime}} 
\mathbb{E}_{\ket{\pi}}\left[ \left|\pi_{i} - \pi_{j}\right| \left|\pi_{i^{\prime}} - \pi_{j^{\prime}}\right| \right], \notag\\
&= \frac{1}{\beta^{2}_{1}} \Biggl(
\sum_{i<j} A^{2}_{ij} \mathbb{E}_{\ket{\pi}}\left[ \left|\pi_{i} - \pi_{j}\right|^{2} \right] \notag\\
&\hspace{30pt} 
+ \sum_{\substack{i,j,k \\ (i<j<k)}} A_{ij} A_{jk} \mathbb{E}_{\ket{\pi}}\left[ \left|\pi_{i} - \pi_{j}\right| \left|\pi_{j} - \pi_{k}\right| \right] 
+ \sum_{\substack{i,j,k \\ (k<i<j)}} A_{ij} A_{ki} \mathbb{E}_{\ket{\pi}}\left[ \left|\pi_{i} - \pi_{j}\right| \left|\pi_{i} - \pi_{k}\right| \right] 
\notag\\
&\hspace{30pt} 
+ \sum_{\substack{i,j,k \\ (i<j, \, i<k, j\ne k)}} A_{ij} A_{ik} \mathbb{E}_{\ket{\pi}}\left[ \left|\pi_{i} - \pi_{j}\right| \left|\pi_{i} - \pi_{k}\right| \right] 
+ \sum_{\substack{i,j,k \\ (i<j, \, k<j, i\ne k)}} A_{ij} A_{kj} \mathbb{E}_{\ket{\pi}}\left[ \left|\pi_{i} - \pi_{j}\right| \left|\pi_{j} - \pi_{k}\right| \right] \notag\\
&\hspace{30pt} 
+ \sum_{\substack{i, j, k, \ell  \\ (i < j < k < \ell)}} A_{ij} A_{k\ell} 
\mathbb{E}_{\ket{\pi}}\left[ \left|\pi_{i} - \pi_{j}\right| \left|\pi_{k} - \pi_{\ell}\right| \right] 
+ \sum_{\substack{i, j, k, \ell  \\ (i < k < j < \ell)}} A_{ij} A_{k\ell} 
\mathbb{E}_{\ket{\pi}}\left[ \left|\pi_{i} - \pi_{j}\right| \left|\pi_{k} - \pi_{\ell}\right| \right] \notag\\
&\hspace{30pt} 
+ \sum_{\substack{i, j, k, \ell  \\ (k < i < j < \ell)}} A_{ij} A_{k\ell} 
\mathbb{E}_{\ket{\pi}}\left[ \left|\pi_{i} - \pi_{j}\right| \left|\pi_{k} - \pi_{\ell}\right| \right] 
+ \sum_{\substack{i, j, k, \ell  \\ (i < k < \ell < j)}} A_{ij} A_{k\ell} 
\mathbb{E}_{\ket{\pi}}\left[ \left|\pi_{i} - \pi_{j}\right| \left|\pi_{k} - \pi_{\ell}\right| \right] \notag\\
&\hspace{30pt} 
+ \sum_{\substack{i, j, k, \ell  \\ (k < i < \ell < j)}} A_{ij} A_{k\ell} 
\mathbb{E}_{\ket{\pi}}\left[ \left|\pi_{i} - \pi_{j}\right| \left|\pi_{k} - \pi_{\ell}\right| \right] 
+ \sum_{\substack{i, j, k, \ell  \\ (k < \ell < i < j)}} A_{ij} A_{k\ell} 
\mathbb{E}_{\ket{\pi}}\left[ \left|\pi_{i} - \pi_{j}\right| \left|\pi_{k} - \pi_{\ell}\right| \right]
\Biggr). \label{SecondMomentH1Decomposition}
\end{align}
Note that $i$, $j$, $k$, and $\ell$ are the raw indices of the vertices that are used to identify the vertices themselves. 
The first term represents the case in which the vertex pairs $(i, j)$ and $(i^{\prime}, j^{\prime})$ are identical. 
The second to fifth terms represent the cases where one of the vertices in $(i, j)$ is identical to one of $(i^{\prime}, j^{\prime})$: 
$j = i^{\prime}$ (third term), 
$i = j^{\prime}$ (fourth term), 
$i = i^{\prime}$ (fifth term), 
and $j = j^{\prime}$ (sixth term).
Finally, the sixth to eleventh terms represent the cases where the vertices for $i$, $j$, $k$, and $\ell$ do not coincide at all. 

The first term is 
\begin{align}
\sum_{i<j} A^{2}_{ij} \mathbb{E}_{\ket{\pi}}\left[ \left|\pi_{i} - \pi_{j}\right|^{2} \right] 
= \frac{1}{\binom{N}{2}} \sum_{a < b} (a - b)^{2} \sum_{i<j} A^{2}_{ij} 
= \frac{M N(N+1)}{6}.
\end{align}

To calculate the sum of the second to fifth terms in Eq.~(\ref{SecondMomentH1Decomposition}), we first calculate the expectation with respect to $\ket{\pi}$: 
\begin{align}
\mathbb{E}_{\ket{\pi}}\left[ \left|\pi_{i} - \pi_{j}\right| \left|\pi_{j} - \pi_{k}\right| \right] 
&= \frac{1}{3! \binom{N}{3}} \Biggl(
2 \sum_{\pi_{i}=1}^{N-2} \sum_{\pi_{j}=\pi_{i}+1}^{N-1} \sum_{\pi_{k}=\pi_{j}+1}^{N} 
\left|\pi_{i} - \pi_{j}\right| \left|\pi_{j} - \pi_{k}\right| \notag\\
&\hspace{40pt}+ 2 \sum_{\pi_{i}=1}^{N-2} \sum_{\pi_{k}=\pi_{i}+1}^{N-1} \sum_{\pi_{j}=\pi_{k}+1}^{N} 
\left|\pi_{i} - \pi_{j}\right| \left|\pi_{j} - \pi_{k}\right| \notag\\
&\hspace{40pt}+ 2 \sum_{\pi_{j}=1}^{N-2} \sum_{\pi_{i}=\pi_{j}+1}^{N-1} \sum_{\pi_{k}=\pi_{i}+1}^{N} 
\left|\pi_{i} - \pi_{j}\right| \left|\pi_{j} - \pi_{k}\right|
\Biggr) \notag\\
&= \frac{(N+1) (7N+4)}{60} \label{SigmaExpectationTripleH1-Appendix}
\end{align}

Note that the third term in Eq.~(\ref{SecondMomentH1Decomposition}) becomes identical to the second term by relabeling the vertex indices as $k \to i$, $i \to j$, and $j \to k$. 
Similarly, we can show that the fourth and fifth terms are identical, using the fact that we can replace $A_{ij}$ with $A_{ji}$ and $A_{ik}$ with $A_{ki}$ in undirected graphs and relabel the vertex indices as $i \to j$ and $j \to i$. 
Therefore, the sum of the second to fifth terms in Eq.~(\ref{SecondMomentH1Decomposition}) reads 
\begin{align}
& 2 \sum_{\substack{i,j,k \\ (i<j<k)}} A_{ij} A_{jk} \mathbb{E}_{\ket{\pi}}\left[ \left|\pi_{i} - \pi_{j}\right| \left|\pi_{j} - \pi_{k}\right| \right] 
+ 2 \sum_{\substack{i,j,k \\ (i<j, \, k<j, i\ne k)}} A_{ij} A_{kj} \mathbb{E}_{\ket{\pi}}\left[ \left|\pi_{i} - \pi_{j}\right| \left|\pi_{j} - \pi_{k}\right| \right] 
= \frac{M_{3} (N+1) (7N+4)}{30},
\end{align}
where $M_{3}$ is the total number of connected edge pairs, or wedges: 
\begin{align}
M_{3} \equiv 
\sum_{\substack{i,j,k \\ (i<j<k)}} A_{ij} A_{jk} 
+ \sum_{\substack{i,j,k \\ (i<j, \, k<j, i\ne k)}} A_{ij} A_{kj}. 
\end{align}

We can analogously calculate the sixth to eleventh terms in  Eq.~(\ref{SecondMomentH1Decomposition}). 
The expectation with respect to $\ket{\pi}$ in each case is 
\begin{align}
\mathbb{E}_{\ket{\pi}}\left[ \left|\pi_{i} - \pi_{j}\right| \left|\pi_{k} - \pi_{\ell}\right| \right] 
&= \frac{1}{4! \binom{N}{4}} \Biggl( 4 
\sum_{\pi_{i}=1}^{N-3} \sum_{\pi_{j} = \pi_{i}+1}^{N-2} \sum_{\pi_{k}=\pi_{j}+1}^{N-1} \sum_{\pi_{\ell}=\pi_{k}+1}^{N} 
\left|\pi_{i} - \pi_{j}\right| \left|\pi_{j} - \pi_{k}\right| \notag\\
&\hspace{40pt}+ 4 
\sum_{\pi_{i}=1}^{N-3} \sum_{\pi_{k} = \pi_{i}+1}^{N-2} \sum_{\pi_{j}=\pi_{k}+1}^{N-1} \sum_{\pi_{\ell}=\pi_{j}+1}^{N} 
\left|\pi_{i} - \pi_{j}\right| \left|\pi_{j} - \pi_{k}\right| \notag\\
&\hspace{40pt}+ 4 
\sum_{\pi_{k}=1}^{N-3} \sum_{\pi_{i} = \pi_{k}+1}^{N-2} \sum_{\pi_{j}=\pi_{i}+1}^{N-1} \sum_{\pi_{\ell}=\pi_{j}+1}^{N} 
\left|\pi_{i} - \pi_{j}\right| \left|\pi_{j} - \pi_{k}\right| \notag\\
&\hspace{40pt}+ 4 
\sum_{\pi_{i}=1}^{N-3} \sum_{\pi_{k} = \pi_{i}+1}^{N-2} \sum_{\pi_{\ell}=\pi_{k}+1}^{N-1} \sum_{\pi_{j}=\pi_{\ell}+1}^{N} 
\left|\pi_{i} - \pi_{j}\right| \left|\pi_{j} - \pi_{k}\right| \notag\\
&\hspace{40pt}+ 4 
\sum_{\pi_{k}=1}^{N-3} \sum_{\pi_{i} = \pi_{k}+1}^{N-2} \sum_{\pi_{\ell}=\pi_{i}+1}^{N-1} \sum_{\pi_{j}=\pi_{\ell}+1}^{N} 
\left|\pi_{i} - \pi_{j}\right| \left|\pi_{j} - \pi_{k}\right| \notag\\
&\hspace{40pt}+ 4 
\sum_{\pi_{k}=1}^{N-3} \sum_{\pi_{\ell} = \pi_{k}+1}^{N-2} \sum_{\pi_{i}=\pi_{\ell}+1}^{N-1} \sum_{\pi_{j}=\pi_{i}+1}^{N} 
\left|\pi_{i} - \pi_{j}\right| \left|\pi_{j} - \pi_{k}\right| 
\Biggr) \notag\\
&= \frac{(N+1)(5N+4)}{45}. \label{SigmaExpectationQuadrupleH1-Appendix}
\end{align}

Therefore, the sum of the sixth to eleventh terms is $2 M_{4} (N+1)(5N+4)/45$, where $M_{4}$ is the total number of disconnected edge pairs: 
\begin{align}
2 M_{4} = 
&\sum_{\substack{i, j, k, \ell  \\ (i < j < k < \ell)}} A_{ij} A_{k\ell} 
+ \sum_{\substack{i, j, k, \ell  \\ (i < k < j < \ell)}} A_{ij} A_{k\ell} 
+ \sum_{\substack{i, j, k, \ell  \\ (k < i < j < \ell)}} A_{ij} A_{k\ell} \notag\\
&+ \sum_{\substack{i, j, k, \ell  \\ (i < k < \ell < j)}} A_{ij} A_{k\ell} 
+ \sum_{\substack{i, j, k, \ell  \\ (k < i < \ell < j)}} A_{ij} A_{k\ell} 
+ \sum_{\substack{i, j, k, \ell  \\ (k < \ell < i < j)}} A_{ij} A_{k\ell}. 
\end{align}

In summary, we have 
\begin{align}
\mathbb{E}_{\ket{\pi}}\left[ \stat{H}^{2}_{1}(\mat{A}, \ket{\pi}) \right] 
&= \left(\frac{3}{M (N+1)}\right)^{2} 
\left(
\frac{M N(N+1)}{6} 
+ \frac{M_{3} (N+1) (7N+4)}{30}
+ \frac{2 M_{4} (N+1)(5N+4)}{45}
\right) \notag\\
&= \left(\frac{3}{M (N+1)}\right)^{2} 
\Biggl(
\frac{M N(N+1)}{6} 
+ \frac{M_{3} (N+1) (7N+4)}{30}
+ \frac{(M^{2} - M - 2M_{3}) (N+1)(5N+4)}{45}
\Biggr) \notag\\
&= 1 + \frac{1}{2M} \left(
\frac{5N-8}{5(N+1)} 
+ \frac{M_{3} (N-4)}{5M (N+1)}
- \frac{2M}{5 (N+1)}
\right)
\end{align}
as the exact solution of the second moment. 
Here, we used the fact that $M_{4} = \binom{M}{2} - M_{3}$ by definition. 
Therefore, the variance of the standardized statistic is 
\begin{align}
\mathrm{Var}_{\ket{\pi}}\left[ 
z_{1}(\mat{A}, \ket{\pi})
\right] 
= \frac{2M (N+1)}{N-2} \mathrm{Var}_{\ket{\pi}} \left[ \stat{H}_{1}(\mat{A}, \ket{\pi}) \right] 
= \frac{N+1}{N-2}  \left(
\frac{5N-8}{5(N+1)} 
+ \frac{M_{3} (N-4)}{5M (N+1)}
- \frac{2M}{5 (N+1)}
\right). 
\label{W1-Appendix}
\end{align}

\end{widetext}

\section{Sequential locality with other affinity metrics \label{sec:HGstatistic}}
We focused on the $\stat{H}_{1}$ test statistic, which uses the sequential distance as an affinity metric, although we could employ other affinity metrics. 
Let us consider sequential locality with a logarithmic semimetric as an example. 
We denote the test statistic as $\stat{H}_{G}$: 
\begin{align}
\stat{H}_{G}(\mat{A}, \ket{\pi}) = -\frac{1}{\beta_{G}} \sum_{i<j} A_{ij} \log\left( 1 - \frac{\left| \pi_{i} - \pi_{j} \right|}{N} \right), \label{HG_Definition}
\end{align}
where $\beta_{G}$ is a normalization constant. 

Analogous to the case of $\stat{H}_{1}$, we consider the following random variable for $\stat{H}_{G}(\mat{A}, \ket{\pi})$: 
\begin{align}
\mathsf{H}_{G} = -\frac{1}{\beta_{G}} \sum_{m=1}^{M} \log\left( 1 - \frac{\mathsf{X}_{m}}{N} \right). 
\end{align}
$\mathsf{X}_{m} \in \mathbb{N}$ is again a random positive integer that independently obeys the discrete triangular distribution. 
By applying the central limit theorem, $\mathsf{H}_{G}$ asymptotically obeys the following distribution when $M$ is sufficiently large: 
\begin{align}
\mathrm{Prob}\left[ \frac{\sqrt{M}}{\sigma_{N}}\left(\frac{\beta_{G} \mathsf{H}_{G}}{M} - \mu_{N} \right) \le a \right] 
= \int_{-\infty}^{a} \frac{dx}{\sqrt{2\pi}} e^{-\frac{1}{2}x^{2}}, \label{HGNullDistributionNormal0}
\end{align}
where $\mu_{N}$ and $\sigma^{2}_{N}$ are the mean and variance of the random variable $-\log\left( 1 - \mathsf{X}/N \right)$, 
\begin{align}
 \mu_{N} 
&= \mathbb{E}_{\mathsf{X}}\left[ -\log\left( 1 - \frac{\mathsf{X}}{N} \right) \right] \notag\\
&= -\sum_{k=1}^{N-1} \log \left( 1 - \frac{k}{N} \right) \frac{2 (N-k)}{N(N-1)} \notag\\
&= \log N - \frac{2}{N(N-1)} \sum_{k=1}^{N-1} k \log k, \label{muN-Appendix}\\
\sigma^{2}_{N} 
&= \mathbb{E}_{\mathsf{X}}\left[ \left( \log\left( 1 - \frac{\mathsf{X}}{N} \right) \right)^{2} \right] - \mu^{2}_{N} \notag\\
&= \frac{2}{N(N-1)} \left( \sum_{k=1}^{N-1} k (\log k)^{2} -2 \log N \sum_{k=1}^{N-1} k \log k \right) \notag\\
&\hspace{10pt}+ (\log N)^{2} 
- \left( \log N - \frac{2}{N(N-1)} \sum_{k=1}^{N-1} k \log k \right)^{2} \notag\\
&= \frac{2}{N(N-1)} \sum_{k=1}^{N-1} k (\log k)^{2} \notag\\
&\hspace{10pt}- \left( \frac{2}{N(N-1)} \sum_{k=1}^{N-1} k \log k \right)^{2}, \label{sigmaN-Appendix}
\end{align}
respectively. 
By setting $\beta_{G} = \mu_{N}M$, we have 
\begin{align}
\mathrm{Prob}\left[ \frac{\mu_{N}}{\sigma_{N}}\sqrt{M} \left(\mathsf{H}_{G} - 1 \right) \le a \right] 
= \int_{-\infty}^{a} \frac{dx}{\sqrt{2\pi}} e^{-\frac{1}{2}x^{2}}. \label{HGNullDistributionNormal}
\end{align}
Hence, 
\begin{align}
z_{G}(\mat{A}, \ket{\pi}) = \frac{\mu_{N}}{\sigma_{N}}\sqrt{M}(\stat{H}_{G}(\mat{A}, \ket{\pi}) - 1)
\end{align}
is the z-statistic for $\stat{H}_{G}(\mat{A}, \ket{\pi})$. 

\begin{figure}[t!]
  \centering
  \includegraphics[width= \columnwidth]{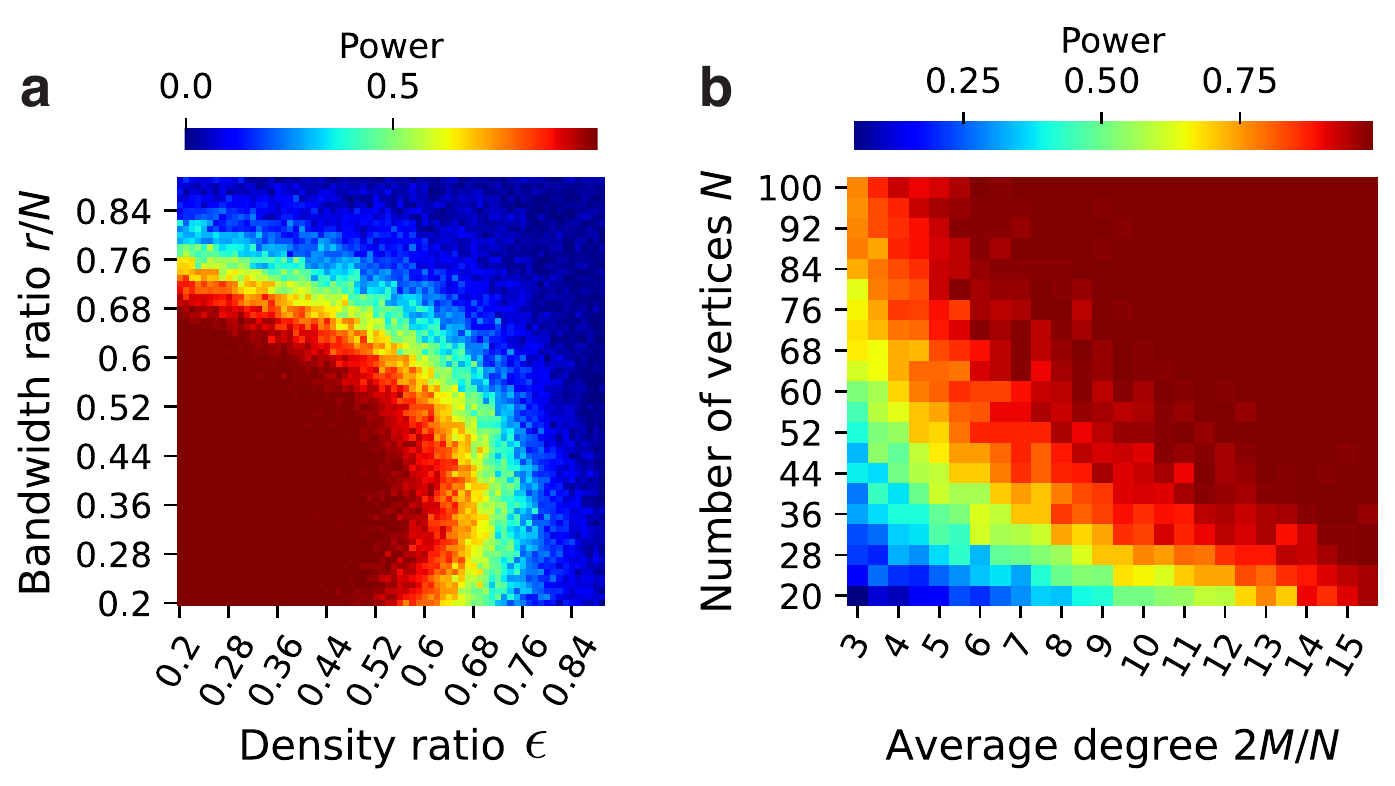}
  \caption{
  Power of the test for unoptimized sequences using the $\stat{H}_{G}$ statistic. 
  While the null hypothesis is the {\ER} model, the graphs are generated by the ORGM. 
  The density plots represent the empirical power, the true-positive rate out of $100$ samples, in (a) the $(r/N, \epsilon)$-plane ($N=50$, $M=200$)  and (b) the $(N, 2M/N)$-plane ($r/N = 0.75$, $\epsilon = 0$). 
	}
  \label{fig:HG_Power_ER_vs_ORGM}
\end{figure}

In Fig.~\ref{fig:HG_Power_ER_vs_ORGM}, we show the power of the $\stat{H}_{G}$ statistic when the graphs are generated by the ORGM. 
This is the same analysis as that for the $\stat{H}_{1}$ statistic in the main text, except that we compute the empirical power based on $100$ ORGM samples at each parameter set, instead of the analytical estimate using the normal approximation. 
In both the $(r/N, \epsilon)$- and $(N, 2M/N)$-planes, the power of the $\stat{H}_{G}$ statistic is higher than that obtained for the $\stat{H}_{1}$ statistic (the red regions in Fig.~\ref{fig:HG_Power_ER_vs_ORGM} are wider). 

We can continue by carrying out the same argument for the $\stat{H}_{1}$ statistic in other parts of this paper. 
We can calculate the first and second moments of $z_{G}$ under the random sequences and show that the variance $\mathrm{Var}_{\ket{\pi}}\left[ z_{G}(\mat{A}, \ket{\pi}) \right]$ is again a function that depends only on $N$, $M$, and $M_{3}$. 
We can also calculate moments under the ORGM. 
However, these results are not written in a compact form because the summations of logarithms cannot be simplified. 
Computing these moments is also inefficient because they require many for-loops to execute the summations. 
Therefore, even though the $\stat{H}_{G}$ statistic has higher power in terms of graph size, we conclude that there is no clear benefit of employing $\stat{H}_{G}$ as the test statistic for the present statistical test.



%

\section*{Acknowledgements}
The authors acknowledge the financial support from JSPS KAKENHI\ 19H01506 (Kawamoto and Kobayashi) and 20H05633 and 22H00827 (Kobayashi). 

%



\end{document}